\documentclass[11pt]{article}

\pdfoutput=1

\usepackage{jheppub}
\usepackage{amsfonts,amsmath,amssymb}
\usepackage{array}
\usepackage{epsfig}
\usepackage{latexsym}
\usepackage{mathtools}
\usepackage{url,bm}

\def\la{{\langle}}
\def\ra{{\rangle}}
\def\del{{\partial}}
\def\ua{{\underline{\alpha}}}
\def\ub{{\underline{\beta}}}
\def\delbar{{\bar\partial}}
\def\eps{{\varepsilon}}
\def\can{{K_\Sigma}}
\def\rD{{\mathrm{D}}}
\def\PT{{\mathbb{PT}}}

\def\v{{\eurm v}}

\def\a{\text{{\teneurm a}}}
\def\b{\text{{\teneurm b}}}
\def\c{\text{{\teneurm c}}}
\def\dd{\text{{\teneurm d}}}
\def\g{\text{{\teneurm g}}}

\def\k{\text{{\teneurm k}}}
\def\m{\text{{\teneurm m}}}
\def\n{\text{{\teneurm n}}}
\def\s{\text{{\teneurm s}}}

\def\sdd{\text{{\eighteurm d}}}
\def\sg{\text{{\eighteurm g}}}
\def\sh{\text{{\eighteurm h}}}
\def\sk{\text{{\eighteurm k}}}
\def\sm{\text{{\eighteurm m}}}
\def\ssd{\text{{\seveneurm d}}}
\def\ssk{\text{{\seveneurm k}}}

\def\Ber{{\mathrm{Ber}}}
\def\rX{{\mathrm{X}}}

\def\be{\begin{equation}}
\def\ee{\end{equation}}

\def\hat{\widehat}
\def\tilde{\widetilde}

\def\sn{\text{{\eighteurm n}}}
\def\D{{\mathcal D}}
\def\YY{{\eusm Y}}
\def\NN{{\eusm N}}
\def\V{{\mathcal V}}
\def\O{{\mathcal O}}
\def\gh{{\mathrm{gh}}}

\def\d{{\mathrm d}}
\def\rZ{{\mathrm Z}}
\def\C{{\mathbb C}}
\def\HH{{\mathbb H}}
\def\CP{{\mathbb{CP}}}
\def\N{{\mathcal N}}
\def\F{{\mathcal F}}

\def\Pi{{\varPi}}
\def\L{{\mathcal  L}}
\def\V{{\mathcal V}}
\def\I{{\mathcal I}}
\def\G{{\mathcal G}}

\def\M{{\mathcal M}}
\def\H{{\mathcal H}}
\def\cZ{\mathcal Z}
\def\tilde{\widetilde}
\def\bar{\overline}

\font\teneurm=eurm10 \font\seveneurm=eurm7 \font\eighteurm=eurm8 \font\fiveeurm=eurm5
\newfam\eurmfam
\textfont\eurmfam=\teneurm \scriptfont\eurmfam=\seveneurm
\scriptscriptfont\eurmfam=\fiveeurm
\def\eurm#1{{\fam\eurmfam\relax#1}}
\font\teneusm=eusm10 \font\seveneusm=eusm7 \font\fiveeusm=eusm5
\newfam\eusmfam
\textfont\eusmfam=\teneusm \scriptfont\eusmfam=\seveneusm
\scriptscriptfont\eusmfam=\fiveeusm
\def\eusm#1{{\fam\eusmfam\relax#1}}
\font\tencmmib=cmmib10 \skewchar\tencmmib='177
\font\sevencmmib=cmmib7 \skewchar\sevencmmib='177
\font\fivecmmib=cmmib5 \skewchar\fivecmmib='177
\newfam\cmmibfam
\textfont\cmmibfam=\tencmmib \scriptfont\cmmibfam=\sevencmmib
\scriptscriptfont\cmmibfam=\fivecmmib

\title{Twistor Strings for $\N=8$ Supergravity}

\author{David Skinner}
\affiliation{School of Natural Sciences, Institute for Advanced Study,\\ Einstein Drive, Princeton, NJ 08540, USA\medskip}
\affiliation{Department of Applied Mathematics and Theoretical Physics,\\ Wilberforce Road, Cambridge CB3 0WA, UK}

\abstract{This paper presents a worldsheet theory describing holomorphic maps to twistor space with $\N$ fermionic directions. The theory is anomaly free when $\N=8$. Via the 
Penrose transform, the vertex operators correspond to an $\N=8$ Einstein supergravity multiplet. In the first instance, the theory describes gauged supergravity in AdS$_4$. 
Upon taking the flat space, ungauged limit, the complete classical S-matrix is recovered from worldsheet correlation functions.}

\begin{document} 

\maketitle


\section{Introduction}
\label{intro} 


The most influential scattering amplitude in Yang-Mills theory is undoubtedly the Parke-Taylor amplitude~\cite{Parke:1986gb}
\be
	\mathcal{A}_{\sn,0} = \frac{\la i,j\ra^4\, \delta^{4}\!\left(\sum_i p_i\right)}{\la1,2\ra\la2,3\ra\cdots\la\n,1\ra}\ .
\label{ParkeTaylor}
\ee
It describes the tree-level (colour-ordered) scattering of two gluons $i$ and $j$ of negative helicity and $\n-2$ gluons of positive helicity, each of momentum 
$p_i = \lambda_i\tilde\lambda_i$. Its talismanic status rests on two pillars. First, its extraordinary simplicity assures that scattering amplitudes are far more managable objects than 
could be expected from momentum space Feynman diagrams, encouraging us that their structures and properties will repay our close attention. Second, it provides a remarkably 
fertile base for deeper explorations of the full Yang-Mills S-matrix. In it, one already sees hints of the twistor action for Yang-Mills~\cite{Mason:2005zm,Boels:2006ir} and the 
associated MHV diagram formalism~\cite{Cachazo:2004kj,Brandhuber:2004yw,Elvang:2008vz,Adamo:2011cb}, the germ of the Grassmannian formulation of all on-shell 
diagrams~\cite{ArkaniHamed:2009dn,Mason:2009qx,ArkaniHamed:2012nw}, and the amplitude's factorization properties --- a crucial ingredient of BCFW 
recursion~\cite{Britto:2005fq} --- laid bare. Finally, and of particular relevance to the present paper, \eqref{ParkeTaylor} is an avatar of Witten's 
representation~\cite{Witten:2003nn}\footnote{Throughout this paper, $(ij)$ denotes the SL$(2;\C)$-invariant inner product $\epsilon_{\ua\ub}\sigma_i^\ua\sigma_j^\ub$  of the 
homogeneous coordinates $\sigma^\ua$ on an abstract curve $\Sigma$ of genus zero. $Z$ denotes a holomorphic map $Z:\Sigma \to \PT$ to $\N$-extended supertwistor space, 
here with $\N=4$. In~\eqref{N=4} this map has degree $\sk+1$. $A_i(Z)$ are twistor representatives of the external wavefunctions. See~\cite{Witten:2003nn} for further details.}
\be
	\mathcal{A}_{\sn,\sk} = \int\frac{\d^{4(\sk+2)|4(\sk+2)}Z}{\rm vol(GL(2;\C))}\,\frac{1}{(12)(23)\cdots(\n 1)}\, \prod_{i=1}^\sn A_i(Z(\sigma_i))\,(\sigma_i\d\sigma_i)
\label{N=4}
\ee
of the $\n$-particle N$^\sk$MHV amplitude in $\N=4$ SYM as an integral over the space of degree $\k+1$ rational curves in twistor space. Witten obtained this form 
by generalizing Nair's interpretation~\cite{Nair:1988bq} of the Parke-Taylor amplitude in terms of (the leading trace part of) a current correlator supported on a holomorphic twistor 
line. 

\medskip

This paper is concerned not with Yang-Mills theory, but with gravity. An expression for all $\n$-particle tree-level MHV amplitudes in gravity was found by Berends, Giele \& 
Kuijf~\cite{Berends:1988zp} only two years after the discovery of the Parke-Taylor amplitude. Despite this, the gravitational S-matrix has proved more resistant to study than its 
Yang-Mills counterpart. Although gravity amplitudes admit a BCFW expansion~\cite{Cachazo:2005ca,Bedford:2005yy,Benincasa:2007qj,ArkaniHamed:2008gz}, actually carrying 
it out leads to expressions for $\n$-particle N$^\sk$MHV amplitudes whose structure is as yet unclear~\cite{Drummond:2009ge}. Applying Risager's procedure to the BGK 
amplitude~\cite{Risager:2005vk,BjerrumBohr:2005jr} leads to an MHV diagram formulation that fails when $\n\geq12$~\cite{Bianchi:2008pu,Conde:2012ik}, while the twistor 
action for gravity tentatively proposed in~\cite{Mason:2008jy} does not appear to extend naturally to $\N=8$ supergravity. Considering that $\N=8$ supergravity is supposed to be 
the simplest quantum field theory, it has been remarkably difficult to grapple with.

The situation changed dramatically with Hodges' two papers~\cite{Hodges:2011wm,Hodges:2012ym}. Hodges showed that the tree-level MHV amplitude for gravity could be 
reformulated as
\be
	\mathcal{M}_{\sn,0} = \la i,j\ra^8\,{\det}'({\rm H})\,\delta^4\!\left(\sum_i p_i\right)\ ,
\label{momHodges}
\ee
where H is the $\n\times\n$ symmetric matrix with entries
\be
	{\rm H}_{ij} = \frac{[i,j]}{\la i,j\ra} \qquad\hbox{for $i\neq j$,} \qquad\qquad {\rm H}_{ii} = -\sum_{j\neq i} \frac{[i,j]}{\la i,j\ra}\frac{\la a,j\ra\la b,j\ra}{\la a,i\ra\la b,i\ra}\ ,
\label{momHodges2}
\ee
where $|a\ra$ and $|b\ra$ are arbitrary spinors. The diagonal entries are nothing but the characteristic gravitational `soft factors' for the $i^{\rm th}$ 
particle~\cite{Weinberg:1965nx,Berends:1988zp,Bern:1998sv}.  H has rank $\n-3$, and ${\det}'(\rm H)$ is any $(\n-3)\times(\n-3)$ minor of H, divided by the permutation 
symmetric combination $\la r_1,r_2\ra\la r_2,r_3\ra\la r_3,r_1\ra$ corresponding to the removed rows and also by a similar factor $\la c_1,c_2\ra\la c_2,c_3\ra\la c_3,c_1\ra$ for 
the removed columns. Hodges' representation has many remarkable properties. Chief among these is that Bose symmetry in the external states is achieved through determinant 
identities rather than through an explicit sum over permutations.

Like the Parke-Taylor amplitude~\eqref{ParkeTaylor}, \eqref{momHodges} provides an inspirational starting point from which to launch deeper
investigations of the gravitational S-matrix. It opens up a path by which to approach gravity on its own terms. In particular, unlike the BGK form, \eqref{momHodges} makes no 
mention of any cyclic ordering that is an artifact of trying to fit gravitational pegs into a Yang-Mills hole. See~\cite{Bern:1998sv,Nguyen:2009jk} for 
deconstructed forms of the Hodges amplitude that were known previously, and~\cite{Feng:2012sy,Adamo:2012xe} for a graph-theoretic explanation of the relation between them.

 \medskip
 
One outcome of these investigations was given in~\cite{Cachazo:2012kg}, where it was conjectured that arbitrary $\n$-particle N$^\sk$MHV tree-level amplitudes in $\N=8$ 
supergravity could be represented as
\be
	\M_{\sn,\sk} = \int\frac{\d^{4(\sk+2)|8(\sk+2)}Z}{\rm vol(GL(2;\C))} \, {\det}'(\HH)\,{\det}'(\HH^\vee)\,\prod_{i=1}^\sn  h_i(Z(\sigma_i))\,(\sigma_i\d\sigma_i)\ .
\label{Freddy}
\ee
This form was obtained by interpreting~\eqref{momHodges} in terms degree 1 holomorphic maps from a Riemann sphere $\Sigma$ into twistor space, and then generalizing 
to higher degree maps. Thus it bears the same relation to~\eqref{momHodges} for $\N=8$ supergravity as~\eqref{N=4} does to~\eqref{ParkeTaylor} for $\N=4$ SYM.  
In~\eqref{Freddy}, $\HH$ is the $\n\times\n$ matrix of operators
\be
	\HH_{ij} = \frac{1}{(ij)}\left[\frac{\del}{\del\mu_i},\frac{\del}{\del\mu_j}\right]\qquad\hbox{for $i\neq j$,}\qquad\qquad
	\HH_{ii} = -\sum_{j\neq i} \frac{1}{(ij)}\left[\frac{\del}{\del\mu_i},\frac{\del}{\del\mu_j}\right]\prod_{r=1}^{\sk+2} \frac{(a_r j)}{(a_r i)}
\label{Hw}
\ee
that act on the external wavefunctions $h_i(Z)$, generalizing~\eqref{momHodges2}. Here, the $a_r$ are any $\k+2$ points on $\Sigma$.  The factor of ${\det}'(\HH)$ 
in~\eqref{Freddy} is any $(\n-\k-3)\times(\n-\k-3)$ minor of $\HH$, divided by the Vandermonde determinant
\be
	|\sigma_{r_1}\cdots\sigma_{r_{\ssk +3}}|\  \equiv	\prod_{\substack{i<j\\ i,j\in \{{\rm removed}\}}}(r_i r_j) 
\label{Vdmintro1}
\ee
of the worldsheet coordinates corresponding to the removed rows, and a similar factor for the removed columns. The combination ${\det}'(\HH)$ is independent of the choice of 
minor. We shall call $\HH$ `the worldsheet Hodges matrix', or often just `the Hodges matrix'. Similarly, $\HH^\vee$ is the $\n\times\n$ matrix with elements
\be
	\HH^\vee_{lm} = \frac{\la\lambda(\sigma_l),\lambda(\sigma_m)\ra}{(lm)}\qquad\hbox{for $l\neq m$,} \quad\quad
	\HH^\vee_{ll}  = -\sum_{m\neq l}\frac{\la\lambda(\sigma_l),\lambda(\sigma_m)\ra}{(lm)}\,\prod_{s=1}^{\sn-\sk-2}\frac{(a_s m)}{(a_s\, l)}\prod_{k\neq l,m}\frac{(k\,l)}{(km)}\, ,
\label{conjH}
\ee
where the $a_s$ are any $\n-\k-2$ points. The factor of ${\det}'(\HH^\vee)$ in~\eqref{Freddy} is any $(\k+1)\times(\k+1)$ minor of $\HH^\vee$, divided by the Vandermonde 
determinant
\be
	|\sigma_{l_1}\cdots\sigma_{l_{\ssk+1}}|\ \equiv \prod_{\substack{l<m\\ l,m\in \{{\rm remain}\}}}(r_l \,r_m) 
\label{Vdmintro2}
\ee
corresponding to the rows that \emph{remain} in this minor, and again by a similar factor for the remaining columns. Again, though it is not obvious from our current description, 
${\det}'(\HH^\vee)$ is completely permutation symmetric in all $\n$ sites. Under a parity transformation of the amplitude, ${\det}'(\HH)$ and ${\det}'(\HH^\vee)$ are 
exchanged~\cite{Cachazo:2012pz,He:2012er,Bullimore:2012cn} (up to a Vandermonde factor involving all $\n$ points), hence we shall call $\HH^\vee$ the `conjugate Hodges 
matrix'. When $\k=0$, the Vandermonde determinant~\eqref{Vdmintro2} should be taken to be unity and ${\det}'(\HH^\vee)$ itself is almost trivial. This is why it is invisible 
in~\eqref{momHodges}.

The conjecture that~\eqref{Freddy} correctly describes the complete classical S-matrix of $\N=8$ supergravity was proved (to a physicist's level of rigour) 
in~\cite{Cachazo:2012pz}, where it was shown that~\eqref{Freddy} obeys the twistor space form of BCFW 
recursion~\cite{Vergu:2006np,Mason:2009sa,ArkaniHamed:2009si,Skinner:2010cz} at the heart of which is the requirement that the amplitude has the correct behaviour in all 
factorization channels. \eqref{Freddy} has also been shown to possess the correct soft limits~\cite{Bullimore:2012cn}. For preliminary investigations of a Grassmannian 
representation of~\eqref{Freddy}, see~\cite{Cachazo:2012pz,He:2012er}. Using this Grassmannian at $\k=0$, an investigation of the MHV diagram formalism for gravity very 
recently been carried out in~\cite{Penante:2012wd}; excitingly, it has the potential to overcome the limitations of Risager's method. A different (presumably equivalent) 
generalization of Hodges' form to higher degree maps can be found in~\cite{Cachazo:2012da,Penante:2012hq}. 

\medskip

The most striking property of the representation~\eqref{Freddy} is that it exists at all. The unfathomable morass of Feynman diagrams that contribute to an $\n$-particle 
gravitational scattering process miraculously conspires to ensure that the tree amplitude lives on a rational curve in twistor space! At MHV, this fact was originally seen by 
Witten in~\cite{Witten:2003nn} using the BGK form of the amplitude\footnote{The derivatives in the Hodges matrix~\eqref{Hodges} 
are responsible for what was called `derivative of a $\delta$-function support' in~\cite{Witten:2003nn}.}. It was also shown to hold for the 5-particle NMHV amplitude 
in~\cite{Giombi:2004ix}\footnote{The non-trivial statement here is just that the 5-point NMHV amplitude has support on some $\CP^2\subset\PT$. Any five points on a plane define 
a conic.}. The existence of~\eqref{Freddy} means, first and foremost, that \emph{all} gravitational tree amplitudes live on higher degree rational curves in twistor space.

\emph{Why} should the gravitational S-matrix know about these curves? The answer pursued here --- really, the only conceivable answer --- is 
that four-dimensional gravity is a twistor string theory. The purpose of the current paper is to find this twistor string. Specifically, over the course of this paper we shall construct a 
worldsheet theory that localizes on holomorphic maps to $\N=8$ supertwistor space and whose vertex operators correspond via the Penrose transform to a linearized $\N=8$ 
supergravity multiplet. We shall obtain~\eqref{Freddy} from worldsheet correlation functions of this theory at $\g=0$. This work thus provides the theoretical framework in 
which~\eqref{Freddy} should be understood. 
See~\cite{Nair:2005iv,AbouZeid:2006wu,Nair:2007md,Brodel:2009ep,Abe:2009kq,Heckman:2011qt,Heckman:2011qu,Adamo:2012nn,Adamo:2012xe} for earlier attempts to 
understand Einstein gravity in the context of twistor strings.

It is clear that~\eqref{Freddy} possesses a rich and intricate structure. What clues are there to guide us to the underlying theory? The main prompt follows from a simple counting 
that also helped stimulate the discovery of~\eqref{Freddy}. At $\g$ loops, $\n$ particle gravitational scattering amplitudes are proportional to the $\kappa^{\sn+2\sg-2}$, where 
$\kappa$ is the square root of the Newton constant G$_{\rm N}$ and has dimensions of (mass)$^{-1}$. Since the amplitude itself is dimensionless, these dimensions must be 
balanced by kinematic factors. But on twistor space, the only object that fixes a mass scale is the infinity twistor --- an antisymmetric bitwistor whose presence breaks conformal 
invariance. With flat space-time, the infinity twistor appears in two separate guises, corresponding to the $\la\ ,\,\ra$ and $[\ ,\, ]$ brackets familiar from spinor momenta. With 
$\n$ particles at $\g$ loops, the twistor space amplitude needs to contain precisely $\n+2\g-2$ factors of $\la\ ,\,\ra$ and $[\ ,\,]$ in total. Under a parity transformation
$\la\ ,\,\ra$ and $[\ ,\,]$ are exchanged, along with the numbers $\n_{\pm}$ of positive and negative helicity gravitons participating in the scattering process (in the pure gravity 
sector). We deduce that the twistor space amplitude must be proportional to 
\be
\begin{aligned}
	\n_+ + \g -1 &= \n-\k+\g-3&&\hbox{factors of $[\ ,\,]$ and}\\
	\n_- + \g - 1 &= \k + \g +1&&\hbox{factors of $\la\ ,\,\ra$}\ .
\end{aligned}
\label{counting}
\ee
Note that the symmetric choice $(\n+2\g-2)/2$ is not possible since $\n$ may be odd and the integrand is rational. The fact that $\n_+$ and $\n_-$ respectively go with $[\ ,\,]$ and 
$\la\ ,\,\ra$ is a convention fixed by comparison with~\eqref{momHodges}. This dependence is certainly realized in~\eqref{Freddy}, where ${\det}'(\HH)$ is easily seen to be a 
monomial of degree $\n-\k-3$ in $[\ ,\,]$ whereas ${\det}'(\HH^\vee)$ is a monomial of degree $\k+1$ in $\la\,,\,\ra$. At higher loops, \eqref{counting} is compatible with all 
factorization channels of $\g$-loop N$^\sk$MHV amplitudes. 

The key question is to ask what sort of worldsheet objects could be responsible for this behaviour. For $[\ ,\,]$, the answer we find is that $\k-\g+3$ of the vertex 
operators are fixed and do not involve the infinity twistor, whereas the remaining $\n-\k+\g-3$ are integrated and are linear in $[\ ,\,]$. To achieve this, the worldsheet theory will 
involve a field with (generically) $\k-\g+3$ zero modes whose fixing is associated to the Vandermonde factor~\eqref{Vdmintro1} in ${\det}'(\HH)$. The dependence on $[\ ,\,]$ in the 
integrated operators is introduced by the BRST operator responsible for the descent procedure. The infinity twistor in the form $[\ ,\,]$ really endows twistor space with a twisted 
holomorphic Poisson structure, and the BRST operator we arrive at is somewhat reminiscent of those found in Poisson sigma models~\cite{Cattaneo:2001ys,Bonechi:2007ar}. 

The theory also contains a different field with (generically) $\k+\g+1$ zero modes whose fixing provides the Vandermonde determinant~\eqref{Vdmintro2} in ${\det}'(\HH^\vee)$. 
The vertex operators associated to this field have a rather different character that may be motivated as follows. Firstly, notice that the previous integrated and unintegrated vertex 
operators have apparently already used up all the available $\n$ insertion points. \eqref{Freddy} allows us to choose the minors of $\HH$ and $\HH^\vee$ completely 
independently, so there seems to be no compelling reason to place the `additional'  operators preferentially with either type of insertion mentioned in the previous paragraph. 
Secondly, although ${\det}'(\HH^\vee)$ knows about the holomorphic map $Z:\Sigma\to\PT$, it is completely oblivious to the external world. Unlike ${\det}'(\HH)$ which knows 
about the external states through the derivative operators in~\eqref{Hw}, nothing in the definition of ${\det}'(\HH^\vee)$ cares what we choose for the wavefunctions $h_i(Z)$, nor 
even how many particles are being scattered. All this is strongly reminiscent of `picture changing operators' of the RNS superstring. See 
\emph{e.g.}~\cite{Friedan:1985ge,Witten:2012bh} for an introduction to these operators. We shall indeed find picture changing operators in our theory, and inserting $\k+\g+1$ of 
these leads to the requisite dependence on $\la\ ,\,\ra$.

The characterization sketched above may sound worryingly piecemeal. On the contrary, the whole theory flows naturally from a single structure: the worldsheet 
carries a certain exotic twisted supersymmetry. All the required objects fit into geometrically meaningful worldsheet supermultiplets --- properly understood, the theory contains 
only three different fields. The action and BRST operator are as simple as one could wish. 

\bigskip

The outline of the paper is as follows. In section~\ref{X} we describe the worldsheet supermanifold whose fermionic symmetries and moduli lie at the heart of the whole 
construction. The actual worldsheet theory is a relative of Berkovits' formulation~\cite{Berkovits:2004hg} of the original twistor string, and is described in section~\ref{model}. 
(Some readers may prefer to begin with this section.) Here we begin with a description of the worldsheet  fields in section~\ref{matter} and BRST operator in section~\ref{BRST}. 
After a brief diversion, we proceed to show in section~\ref{anomalies} that the model is anomaly free if and only if the target twistor space has  $\N=8$ supersymmetry. We 
conclude our discussion of the general theory in section~\ref{vertex}, presenting the vertex operators of the model and explaining their relation to an $\N=8$ supergravity multiplet. 
Section~\ref{amps} contains the derivation of the complete flat space S-matrix of classical $\N=8$ supergravity~\eqref{Freddy} from correlation functions of vertex operators on the 
worldsheet. The model of section~\ref{model} describes SO(8) gauged supergravity on an AdS$_4$ background in the first instance. Thus, before embarking on the S-matrix 
calculation, in section~\ref{flat} we show in section~\ref{flat} how to rescale the fields so as to take the flat space limit. The Hodges matrix~\eqref{Hw} and the conjugate Hodges 
matrix~\eqref{conjH} have a different origin on the worldsheet. They are obtained in sections~\ref{Hodges} and~\ref{cHodges}, respectively. It is worth pointing out immediately 
that the present model is more successful than the original twistor strings~\cite{Witten:2003nn,Berkovits:2004hg} were (as a theory of pure $\N=4$ SYM) in at least one respect: 
the worldsheet correlator we consider leads inexorably to~\eqref{Freddy} and only to~\eqref{Freddy}. No terms are ignored or discarded by hand. Our work suggests many 
promising avenues for future research. We conclude in section~\ref{discussion} with a brief discussion of some of these.


\section{The worldsheet supermanifold}
\label{X}

In this section we describe the geometry of the worldsheet supermanifold X on which the twistor string theory is based. See 
\emph{e.g.}~\cite{Manin:1988ds,Deligne,Witten:2012ga} for much more information on complex supermanifolds.

\medskip

Let $\Sigma$ be a closed, compact Riemann surface of genus $\g$. We extend $\Sigma$ to a complex supermanifold X of dimension $1|2$ by picking\footnote{Eventually, the 
twistor string path integral will include a sum (or integral) over all such choices.} a line bundle $\L\to\Sigma$ of degree $\d\geq0$ and a choice of spin bundle $K_\Sigma^{1/2}$. 
We then define X to be the split supermanifold whose tangent bundle $T\rX$ is
\be
	T\rX = T\Sigma\oplus \D\ ,
\label{TX}
\ee
where $\D$ is the rank 2 fermionic bundle\footnote{Here, $\Pi$ is the `parity reversing functor' whose r{\^o}le is simply to remind us that $\D$ is fermionic.}
\be
	\D \cong \Pi\left(\C^2\otimes  K_\Sigma^{-1/2}\otimes\L\right)
\label{Ddef}
\ee
over $\Sigma$. We will often say that objects taking values in $K_\Sigma^p\otimes \L^q_{\phantom{\Sigma}}$ have spin $p$ and charge $q$. Thus, sections of $\D$ have spin 
$-\frac{1}{2}$ and charge $+1$.

For a local description of X, we cover the bosonic Riemann surface $\Sigma$ by open coordinate patches $U_\alpha$ and let $\{\hat U_\alpha\}$ be the corresponding cover of X. 
Each such $\hat U_\alpha$ is (an open subset of) $\C^{1|2}$, so we may describe X locally in terms of one bosonic and two fermionic holomorphic coordinates $z|\theta^a$, 
with $a=1,2$. The fact that X is a \emph{split} supermanifold means that on overlaps the coordinate transformations are
\be
\begin{aligned}
	z_\alpha &= f_{\alpha\beta}(z_\beta)\\
	\theta^a_\alpha &  
	= \left(g_{\alpha\beta}(z_\beta)\right)^a_{\ b}\theta_\beta^b ,
\end{aligned}
\label{coordtrans}
\ee 
where the transition functions $f_{\alpha\beta}$ and $g_{\alpha\beta}$ on $\hat U_\alpha\cap \hat U_\beta$ depend only on the bosonic coordinate $z$, not 
$(z|\theta)$. To identify these transition functions, suppose we write an arbitrary section $V:\rX\to T\rX$ of the tangent bundle of X as
\be
	V = V^z(z|\theta)\frac{\del}{\del z} + \V^a(z|\theta)\frac{\del}{\del\theta^a}\ .
\label{sectionTX}
\ee
Recalling that $T\rX\cong T\Sigma\oplus\D$, we see that $V^z\del_z$ is a section of $T\Sigma$ (written in terms of the local basis $\del/\del z$) whilst $\V^a\del_a$ is a section of $\D$ (written in terms of the local basis $\del/\del\theta^a$). In order to compensate the transformations of the basis
\be
	\frac{\del}{\del z_\alpha} = \frac{1}{f'_{\alpha\beta}}\frac{\del}{\del z_\beta}  \qquad\qquad\hbox{and}\qquad\qquad
	\frac{\del}{\del\theta^a_\alpha} = (g_{\alpha\beta}^{-1})_a^{\ b}\frac{\del}{\del\theta^b_\beta}
\ee
that follow from~\eqref{coordtrans}, the components $V^z$ and $\V^a$ must transform as
\be
	V^z(z_\alpha|\theta_\alpha) = f_{\alpha\beta}'\,\hat V^z(z_\beta|\theta_\beta) \qquad\qquad\hbox{and}\qquad\qquad
	\V^a(z_\alpha|\theta_\alpha) = \left(g_{\alpha\beta}\right)^a_{\ b} \hat\V^b(z_\beta|\theta_\beta)
\label{comptrans}
\ee
on overlaps. But since $\D\cong \C^{0|2}\otimes T_\Sigma^{1/2}\otimes\L$, we have
\be
	(g_{\alpha\beta})^a_{\ b} = \sqrt{f'_{\alpha\beta}} \times  (h_{\alpha\beta})^a_{\ b}
\label{gspecific}
\ee
where each component of the $2\times 2$ matrix $h$ is a transition function for sections of $\L$.

Because X is a split supermanifold, it can be viewed as the total space of a bundle over $\Sigma$ --- indeed, this is just what is meant by the transformation 
laws~\eqref{coordtrans}. To identify this bundle, note that by~\eqref{gspecific} and~\eqref{comptrans}, the coordinates $\theta$ themselves transform as components of a section of 
$\D$. But since the coordinates on a bundle transform oppositely to the bundle itself, we find that X is the total space of $\D^\vee\to\Sigma$, where $\D^\vee$ is the dual of $\D$.  
To say this differently, functions on X are superfields $\Phi(z|\theta)$ that may be expanded in the usual way as
\be
	\Phi(z|\theta) = \phi(z) + \theta^a\psi_a(z) + \frac{1}{2}\epsilon_{ab}\theta^a\theta^b \xi(z)
\label{FunX}
\ee
where $\phi(z)$ is a function on $\Sigma$, the $\psi_a$ are a pair of functions on $\Sigma$ with values in $K_\Sigma^{+1/2}\otimes\L^{-1}$ and of opposite Grassmann parity to 
$\phi$, and $\xi$ is a function on $\Sigma$ with values in $K_\Sigma\otimes\L^{-2}$. Thus, the structure sheaf of X is $\O_\rX =\O_\Sigma(\wedge^*\D^\vee)$. More generally, if 
$\Phi^{[p,q]}(z|\theta)$ is a section of $K_\Sigma^p\otimes\L^q_{\phantom{\Sigma}}$, then it may be expanded in terms of fields $\phi^{[p,q]}$, $\psi_a^{[p,q]}$ and $\xi^{[p,q]}$ 
on $\Sigma$, of {(spin, charge)} = $(p,q)$, $(p+\frac{1}{2},q-1)$ and $(p+1,q-2)$, respectively.

\medskip

To give an example that will be important later, suppose that $\Sigma$ is the Riemann sphere. On $\CP^1$, the bundles $K^{-1/2}$ and $\L$ are 
uniquely determined to be $\O(1)$ and $\O(\dd)$, respectively. Thus $\D=\C^{0|2}\otimes\O(\dd+1)$ and the  supermanifold X may be identified as the weighted projective 
superspace $\mathbb{WCP}_{(1,1|\sdd+1,\sdd+1)}$ with homogeneous coordinates $(\sigma^\ua|\vartheta^a)$ obeying the scaling
\be
	(\sigma^\ua|\vartheta^a)\equiv (r\sigma^\ua| r^{\sdd+1}\,\vartheta^a) \qquad\qquad\hbox{for any $r\in\C^*$}\ .
\label{WCP}
\ee
In this case, a function $\Phi\in\O_\rX$ may be expanded as
\be
	\Phi(\sigma|\vartheta) = \phi(\sigma) + \vartheta^a\psi_a(\sigma) + \frac{1}{2}\epsilon_{ab}\vartheta^a\vartheta^b\xi(\sigma)\ ,
\ee
where $\phi$, $\psi_a$ and $\xi$ have homogeneities $0$, $-(\dd+1)$ and $-2(\dd+1)$, respectively  under the scaling~\eqref{WCP}.

\medskip

Returning to the general case, the cotangent bundle $T^\vee\rX$ to the supermanifold is just the direct sum $T^\vee\rX \cong K_\Sigma\oplus\D^\vee$ dual to~\eqref{TX}. Thus 
the holomorphic Berezinian $\Ber(\rX)$ of X is
\be
	\Ber(\rX) = \Ber(K_\Sigma)\otimes \Ber(\D^\vee)\ .
\ee
To compute this, recall that for an even parity (bosonic) bundle $\Ber(B) = {\rm Det}(B)$ --- the top exterior power of $B$. However, for an odd parity (fermionic) bundle 
$\Ber(F)={\rm Det}(\Pi F)^{-1}$, where the power of $-1$ appears as a consequence of the fact that in Berezin integration the integral form $\d\theta_1\d\theta_2$ transforms 
oppositely to the differential form $\d\zeta_1\wedge\d\zeta_2$ involving variables $\zeta_a$ that have same quantum numbers, but opposite Grassmann parity to 
$\theta^a$. (In one dimension, this is just the familiar statement that since $\int \d\theta\, \theta = 1$ by Berezin integration, if $\theta\to g(z)\theta$, we require that the 
integral form $\d\theta\to g^{-1}(z)\d\theta$.) Thus we have
\be
\begin{aligned}
	\Ber(\rX) &= K_\Sigma \otimes {\rm Det}(\Pi\D^\vee)^{-1} = K_\Sigma\otimes\left(K_\Sigma\otimes\L^{-2}\right)^{-1}\\
	&\cong\L^2\ ,
\end{aligned}
\label{Ber}
\ee
where we used the definition~\eqref{Ddef}  in the second step.

When we come to write the worldsheet action in section~\ref{model}, we will need a top holomorphic integral form on X. Since the Berezinian of X is isomorphic to $\L^2$, 
the product $\Ber(\rX)\otimes\L^{-2}$ is trivial. Thus it admits a global holomorphic section that we write as $\d^{1|2}z$. For example, at genus zero
$\d^{1|2}z = (\sigma\d\sigma)\d\vartheta_1\d\vartheta_2$ in terms of the homogeneous coordinates $(\sigma^\ua|\vartheta^a)$ introduced above. We can 
treat $\d^{1|2}z$ as a top holomorphic (integral) form on X of charge $-2$. In order to construct an action, this charge must be balanced by the worldsheet Lagrangian $L$, so that 
$\d^{1|2}z\,L(z|\theta)$ may be integrated over X.

\medskip

Let us close this subsection with a couple of remarks. As usual for complex supermanifolds (and as on the twistor target space $\CP^{3|\N}$) we take X to be a cs 
manifold~\cite{Manin:1988ds,Deligne}, in the sense that the antiholomorphic tangent bundle is $\overline {T\rX} \equiv\overline{T\Sigma}$ and so has rank $1|0$. Antiholomorphic 
fermionic directions simply do not exist --- all operations with the fermions will be purely algebraic. Finally, we note that X is \emph{not} an $\N=2$ super Riemann surface 
(see \emph{e.g.}~\cite{Witten:2012ga}), because our choice of $T\rX$ means the distribution $\D$ is integrable in the sense that $\{\D,\D\}\subseteq \D$. Indeed, the usual 
superderivatives 
\be
	D_1 = \frac{\del}{\del\theta^1} + \theta^2\frac{\del}{\del z} \qquad\qquad D_2 = \frac{\del}{\del\theta^2} + \theta^1 \frac{\del}{\del z}
\ee
on an $\N=2$ super Riemann surface do not make any sense for us, because the second term in each expression has different charge from the first and hence is forbidden. 
Exactly these forbidden terms are responsible for the non-integrability of the odd distribution on an $\N=2$ super Riemann surface.


\subsection{Automorphisms}
\label{aut}

We now consider the symmetries of X as a complex supermanifold. On a local patch $\hat U_\alpha$, as usual these are generated by vector fields 
$V\in \Omega^0(\hat U_\alpha,T\rX|_{\hat U_\alpha})$. We will be particularly interested in the symmetries of the distribution $\D$ --- these are generated by the vector fields 
$\V\in\Omega^0(\hat U_\alpha,\left.\D\right|_{\hat U_\alpha})$ that act trivially on $\Sigma$. (This restriction would not make sense on a super Riemann surface, but does make 
sense on X precisely because $\D$ is integrable.)

From~\eqref{sectionTX} we can write
\be
	\V^a(z|\theta)\frac{\del}{\del\theta^a} = \left(\v^a(z) + \theta^b R^a_{\ b}(z) + \frac{1}{2}\epsilon_{bc}\theta^b\theta^c\, \tilde \v^a(z)\right)\frac{\del}{\del\theta^a}
\label{sectionD}
\ee
where the components $\v,R$ and $\tilde\v$ have (spin, charge) = $(-\frac{1}{2},1)$, $(0,0)$ and $(+\frac{1}{2},-1)$, respectively. Because the fermionic distribution $\D$ is 
integrable, the anticommutator $\{\V_1,\V_2\}$ of any two such vector fields again lies in $\D$. A short calculation shows that the 
component fields obey the algebra
\be
\begin{aligned}
	[\v_1, R_2 ] &= \v_{12} \qquad\qquad &[\tilde\v_1, R_2 ] &= \tilde v_{12}\\
	[R_1,R_2] &= R_{12} \qquad\qquad &\{\v_1,\tilde\v_2\} &= R'_{12}\ ,
\end{aligned}
\label{algebra}
\ee
where
\be
\begin{aligned}
	\v_{12}^a &= (R_2)^a_{\ b}\v_1^b\quad\qquad&\tilde\v_{12}^a&=(R_2)^a_{\ b}\tilde\v_1^b -{\rm tr}(R_2)\tilde\v^a_1\\
	(R_{12})^a_{\ b}&=(R_2R_1-R_1R_2)^a_{\ b}\qquad\qquad&(R'_{12})^a_{\ b} &= -\tilde\v_2^a\,\v_{1b}\ .
\end{aligned}
\label{algebra2}
\ee
with $\v_b=\epsilon_{bc}\v^c$. All other commutators are zero --- in particular, $\{\tilde\v_1,\,\tilde\v_2\}=0$ since it is of order $(\theta)^3$ which must vanish. 

If we decompose the $gl(2;\C)$ matrix $R$ as
\be
	R^a_{\ b} =\frac{1}{2}\delta^a_{\ b} \,r+ r^a_{\ b}
\label{Rsplit}
\ee
where the traceless, symmetric matrix $r^a_{\ b}$ takes values in ${sl}(2;\C)$ while $r={\rm tr}(R)$ takes values in ${gl}(1;\C)$. $R$ may be interpreted as generating a gauge 
transform of $\C^2\otimes\L$ so that $r$ generates gauge transformations associated to the determinant $\L^2$. Equations~\eqref{algebra}-\eqref{algebra2} then reflect the fact 
that the ${\eurm v}^a$ transform in the fundamental representation of SL$(2;\C)$ and have charge $+1$ under $\L$, whereas the $\tilde\v^a$ transform in the fundamental of 
SL$(2;\C)$ but have charge $-1$ under $\L$.

Later, in section~\ref{BRST} we shall introduce a (bosonic) ghost multiplet in  $\Pi\Omega^0(\rX,\D)$ corresponding to~\eqref{sectionD}. This algebra will then be interpreted as 
the gauge algebra of our worldsheet theory. Zero modes of the ghost multiplet live in $H^0(\rX,\D)$, parity reversed, and correspond to globally defined $\Sigma$-preserving 
infinitesimal automorphisms of X as a complex supermanifold.


\subsection{Deformations}
\label{def}

As for a usual complex manifold, infinitesimal deformations of X as a complex supermanifold are parametrized by elements of $H^1(\rX,T\rX)$. This cohomology group is the 
holomorphic tangent space to the moduli space of X. Again, we will be interested in the moduli of X associated to deforming the choice of distribution $\D$ whilst leaving 
$\Sigma$ fixed. Infinitesimally, these are described by  $H^1(\rX,\D)\subset H^1(\rX,T\rX)$. On a supermanifold, the dualizing sheaf is the holomorphic Berezinian, so the 
deformations are Serre dual\footnote{See \emph{e.g.}~\cite{Penkov,Haske,Witten:2012ga} for a discussion of Serre duality for complex supermanifolds.} to 
$H^0(\rX,\Ber(\rX)\otimes \D^\vee)$. We showed in~\eqref{Ber} that $\Ber(\rX)\cong\L^2$, so using~\eqref{Ddef} we can identify this cohomology group as
\be
	H^0(\rX,\Ber(\rX)\otimes \D^\vee)\cong \Pi H^0(\rX, \C^2\otimes K_\Sigma^{1/2}\otimes\L)\ ,
\label{moduli}
\ee
where again the symbol $\Pi$ denotes that the fibres are fermionic. This group is non-trivial provided only $\deg(\L)>0$, so 
X has odd moduli even at $\g=0$. In section~\ref{BRST} we shall introduce a (bosonic) antighost multiplet valued in $\Omega^0(\rX,\C^2\otimes K_\Sigma^{1/2}\otimes\L)$. Zero 
modes of this antighost live in~\eqref{moduli}, parity reversed, and so by Serre duality can be paired with deformations of the odd moduli.  This is the usual mechanism by which 
the RNS superstring provides a top holomorphic integral form on odd moduli space; see \emph{e.g.}~\cite{Witten:2012ga,Witten:2012bh}.


\section{The twistor string}
\label{model}

In this section we define the worldsheet theory that will provide a twistor description of Einstein supergravity. After introducing the fields and explaining their geometric meaning, 
we study the gauge and BRST transformations naturally associated to the structure of the worldsheet supermanifold X. The model is chiral, and we show that all (local) worldsheet 
anomalies vanish if and only if the target space has $\N=8$ supersymmetry. We then construct vertex operators in the BRST cohomology, finding that they correspond to an 
$\N=8$ supergravity multiplet. We have just seen that X has odd moduli even at genus zero. We construct the associated `picture changing' operators.

\subsection{Matter fields}
\label{matter}

To define the worldsheet model, we first introduce four bosonic and $\N$ fermionic fields $\cZ^I$ (where $I = 1,\ldots,4|1,\ldots,\N$).
Each of these are scalars on X of charge $+1$. In other words,
\be
	\cZ\in\Omega^0(\rX,\C^{4|\N}\otimes\L)
\label{Z}
\ee
where $\L$ is the same degree $\d$ line bundle used in the definition~\eqref{Ddef} of $\D$. In the first instance, $\cZ$ defines a smooth map $\cZ: \rX\to \C^{4|4}$. The twisting by 
$\L$ means that this map is defined only up to an overall non-zero complex rescaling, so $\cZ$ really defines a map $\cZ:\Sigma\to\CP^{3|\N}$. The $\cZ^I$ then represent the 
pullbacks to X of homogeneous coordinates on this projective space.

Saying that $\cZ$ is a map from X, rather than from $\Sigma$, simply means that it is a worldsheet superfield. As in~\eqref{FunX}, we define 
its component expansion to be
\be
	\cZ^I(z,\theta) = Z^I(z) + \theta^a\rho^I_a(z)  + \frac{1}{2}\theta^a\theta_a\, Y^I(z)
\label{Zcomps}
\ee
in terms of fields $(Z^I, \rho_a^I, Y^I)$ on $\Sigma$. Since each $\theta^a$ is a fermionic coordinate of (spin, charge) = $(-\frac{1}{2},1)$ we see that $Z^I$, like $\cZ^I$, 
is a scalar on $\Sigma$ of charge $+1$,  $\rho_a^I$ are a pair of uncharged $(\frac{1}{2},0)$-forms, while $Y^I$ is a (1,0)-form of charge $-1$. If the index $I$ denotes a bosonic 
direction in $\CP^{3|\N}$, then $Z^I$ and $Y^I$ are bosons while $\rho^I_a$ are fermions. This is reversed when $I$ denotes a fermionic direction. Altogether, the component 
fields in the matter multiplet are
\be
\begin{aligned}
	Z\ &\in\ \Omega^0(\Sigma,\C^{4|\N}\otimes\L)\\
	\rho\ &\in\ \Pi\Omega^0(\Sigma,\C^{4|\N}\otimes\C^2\otimes K_\Sigma^{1/2})\\
	Y\ &\in\ \Omega^0(\Sigma,\C^{4|\N}\otimes\can\otimes\L^{-1})\ .
\end{aligned}
\ee
Of course, the $Z^I$ represent the pullbacks, now to $\Sigma$, of homogeneous coordinates of $\CP^{3|\N}$. We will sometimes decompose $Z^I$ as $Z^I = ({\rm Z^a}|\chi^A) = (\mu^{\dot\alpha},\lambda_\alpha | \chi^A)$ into its bosonic and fermionic components.


\subsubsection{The infinity twistor}
\label{infinity}

In order to write an action for these fields, we must pick some extra data. This is a choice of constant, graded skew symmetric bi-twistor $\I_{IJ}$, known as the `infinity twistor'.
Graded skew-symmetry means that $\I_{IJ} = -(-1)^{|IJ|}\,\I_{JI}$, where $|IJ| = 1$ if both $I$ and $J$ denote fermionic directions, and zero otherwise. 
For any two twistors $Z_1$ and $Z_2$, we will usually denote $\I_{IJ}Z^I_1Z^J_2$ by $\la Z_1,Z_2\ra$.

Projective twistor space carries a natural action of SL$(4|\N;\C)$ acting as linear transformations on the homogeneous coordinates. This is the 
complexification of (the double cover of) the space-time $\N$-extended superconformal group. The r{\^o}le of the infinity twistor is to break conformal invariance and determine a 
preferred metric on space-time~\cite{Penrose:1976jq,Atiyah:1978wi,Ward:1980am}. Specifically, if $X^{\rm ab} = {\rm Z_1^{[a}Z_2^{b]}}$ are homogeneous coordinates for 
the bosonic part of the twistor line $Z_1Z_2$, representing a point $x$ in space-time, then 
\be
	\d s^2 = \frac{\epsilon_{\rm abcd} \d X^{\rm ab}\d X^{\rm cd}}{(\I_{\rm ef}X^{\rm ef})^2}
\label{Itog}
\ee
is the space-time metric. According to this metric, lines in twistor space that obey $\I\cdot X=0$ lie `at infinity' in space-time. The fermionic-fermionic components $\I_{AB}$ were 
examined in~\cite{Wolf:2007tx,Mason:2007ct} where it was shown that they likewise define a metric on the R-symmetry group. A non-trivial $\I_{AB}$ thus corresponds to 
\emph{gauging} the R-symmetry.

For definiteness, we will make the choice
\be
\I_{IJ} = \left(
			\begin{array}{c|c}
				\I_{\rm ab}\,  & \,0 \\
				\hline
				0\,  & \,\I_{AB}
			\end{array}
		\right)
\ee 
where the even-even components $\I_{\rm ab}$ and odd-odd components $\I_{AB}$ are given by\footnote{The fact that $\I_{IJ}$ with lower indices contains $\epsilon^{\alpha\beta}$ with raised indices originates with our conventions that the bosonic twistor components ${\rm Z^a}$ are written as $(\mu^{\dot\alpha},\lambda_\alpha)$.}
\be
\I_{\rm ab} = \begin{pmatrix}
			\Lambda \epsilon_{\dot\alpha\dot\beta} & 0 \\
			0 & \epsilon^{\alpha\beta}
		  \end{pmatrix}
\qquad\qquad\hbox{and}\qquad\qquad
\I_{AB} = \sqrt{\Lambda}\,\delta_{AB}\ ,
\label{Icomps}
\ee
respectively. This infinity twistor is non-degenerate. Its inverse is $\I^{IJ}/\Lambda$, where
\be
	\I^{IJ} =\left(
			\begin{array}{cc|c}
				\epsilon^{\dot\alpha\dot\beta} & 0 & 0 \\
				0 & \Lambda\epsilon_{\alpha\beta} & 0 \\ 
				\hline
				\phantom{0} & \phantom{0} & \phantom{0} \\ [-1.2em]
				0 & 0 & \sqrt \Lambda\, \delta^{AB}
			\end{array}
		\right)\ .
\label{Poisson}
\ee
$\I^{IJ}$ defines a holomorphic Poisson structure $\I^{IJ}\frac{\del}{\del Z^I}\wedge\frac{\del}{\del Z^J}$ of homogeneity $-2$ on twistor space. It will play an important r{\^o}le in the vertex operators.

In~\eqref{Icomps}-\eqref{Poisson}, $\Lambda$ is a constant of dimensions (mass)$^2$. The powers of $\Lambda$ can be understood as follows. Since $x$ has dimensions 
(mass)$^{-1}$ and $\lambda$ has dimensions (mass)$^{\frac{1}{2}}$, the incidence relations $\mu^{\dot\alpha} = x^{\alpha\dot\alpha}\lambda_\alpha$ show that $\mu$ has 
dimensions (mass)$^{-\frac{1}{2}}$. Similarly, the space-time fermionic coordinate $\vartheta$ has dimensions (mass)$^{-\frac{1}{2}}$, so the twistor space fermionic directions 
$\chi$ are dimensionless. The powers of $\Lambda$ ensure that both
\be
	\I_{IJ}Z^I\d Z^J = \Lambda\epsilon_{\dot\alpha\dot\beta}\mu^{\dot\alpha}\d\mu^{\dot\beta} + \epsilon^{\alpha\beta}\lambda_\alpha\d\lambda_\beta 
				+ \sqrt\Lambda\,\delta_{AB}\,\chi^A\d \chi^B
\ee
and the Poisson structure
\be
	\I^{IJ}\frac{\del}{\del Z^I}\wedge\frac{\del}{\del Z^J} = \epsilon^{\dot\alpha\dot\beta}\frac{\del}{\del\mu^{\dot\alpha}}\wedge\frac{\del}{\del\mu^{\dot\beta}}
											\,+\,\Lambda\epsilon_{\alpha\beta}\frac{\del}{\del\lambda_\alpha}\wedge\frac{\del}{\del\lambda_\beta} 
											\,+ \,\sqrt\Lambda\,\delta^{AB}\frac{\del}{\del \chi^A}\odot\frac{\del}{\del \chi^B}
\ee
have homogeneous dimension (mass)$^{+1}$. Bosonically at least, this dimension is important in ensuring that~\eqref{Itog} indeed has dimensions (mass)$^{-2}$ as expected for a space-time metric. Recall that the $\n$-particle $\g$-loop gravitational scattering amplitude comes with a factor of $\kappa^{2\sg-2+\sn}$. These dimensions must be balanced by a total of $2\g-2+\n$ powers of $\I$.

\medskip

With the choice~\eqref{Icomps}, the incidence relations show that the space-time metric~\eqref{Itog} becomes
\be
	\d s^2 = \frac{\eta_{\mu\nu}\d x^\mu\d x^\nu}{(1+\Lambda x^2)^2}
\label{AdSmetric}
\ee
where $\eta_{\mu\nu}$ is the flat metric. This is the metric of (complexified) AdS$_4$ with cosmological constant $\Lambda$. Similarly, with 
$\I_{AB} = \sqrt\Lambda\,\delta_{AB}$ the SL$(\N;\C)$ R-symmetry is broken to SO$(\N;\C)$.  Thus, with the choice~\eqref{Icomps} of infinity twistor, our model will describe 
(subject to an appropriate reality condition) SO$(\N)$ gauged supergravity on an AdS$_4$ background. It is straightforward to introduce an arbitrary gauge coupling for the 
gauged R-symmetry by rescaling $\I_{AB} \to {\rm g}\sqrt\Lambda\, \delta_{AB}$ for some dimensionless coupling g. In section~\eqref{flat} we shall take the limit $\Lambda\to0$ 
(with g remaining fixed) so as to compute the flat space S-matrix of ungauged supergravity. Until then, we set $\Lambda=1$ and ${\rm g}=1$ so as to lighten the notation.


\subsubsection{The action}
\label{action}

Having chosen our infinity twistor, the action for $\cZ$ is simple to state. We have\footnote{In computing the component expansion of this action, we use the conventions 
$\int \d^2\theta\ \theta^a\theta_a= 2$ where $\theta^a\theta_a= \epsilon_{ab}\theta^a\theta^b$. Note also that $\theta^a\theta^b = -\frac{1}{2}\epsilon^{ab} \,\theta^c\theta_c$.}
\be
\begin{aligned}
	{\rm S}_1  &= \frac{1}{4\pi} \int_\rX \d^{1|2}z\,\la \cZ,\delbar\cZ\ra\\
	&= \frac{1}{2\pi}\int_\Sigma \la Y,\delbar Z\ra - \frac{1}{2}\la \rho^a,\delbar\rho_a\ra\ ,
\end{aligned}
\label{S1}
\ee
where $\rho^{aI}=\epsilon^{ab}\rho_b^I$. Notice that the charge $+2$ of $\la \cZ,\delbar\cZ\ra$ balances the charge $-2$ of $\d^{1|2}z$. In writing this action, we let $\delbar$ 
denote the (covariant) Dolbeault operator acting on sections of the appropriate bundles --- since the (0,2)-part $F^{0,2}$ of the curvature of any bundle vanishes trivially on 
restriction to a Riemann surface, we can always work in a gauge in which the (0,1)-form part of any connection is the usual $\delbar$-operator. (This applies equally to the cs 
manifold X.)

Clearly, to have sensible kinetic terms for all components of $\cZ$, it is important that the infinity twistor $\la\ ,\, \ra$ be totally non-degenerate, both in the bosonic directions and 
fermionic directions. This motivates our choice~\eqref{Icomps}. When we come to take the flat space limit in section~\ref{amps} so as to describe scattering amplitudes, the infinity 
twistor necessarily becomes degenerate. We shall then need to rescale the fields so as to remove the dependence on the cosmological constant from the action, at the cost of 
including it in the definition of the $\cZ(z,\theta)$ supermultiplet. 

\medskip

The $YZ$-system\footnote{Here we have used the infinity twistor $\I$ to lower the index on $Y$, so that $Y_I = \I_{IJ}Y^J$. With a non-degenerate infinity twistor, this is harmless, 
but in the flat space case this operation must be done with care. See section~\ref{flat}.}
\be
	 {\rm S}_{YZ} = \frac{1}{2\pi}\int_\Sigma Y_I\delbar Z^I
\ee
was a key ingredient of Berkovits' twistor string~\cite{Berkovits:2004hg}. Here, as there, performing the path integral over $Y$ will lead to the constraint that $Z$ be 
a \emph{holomorphic} section of $\C^{4|\N}\otimes\L$. Thus, on-shell, $Z$ describes a \emph{holomorphic} map to $\PT$.  Berkovits' twistor string describes non-minimal $\N=4$ 
conformal supergravity~\cite{Fradkin:1985am,Berkovits:2004jj,Dolan:2008gc}, at least at tree-level, as does Witten's original model~\cite{Witten:2003nn}. The $\N=4$ (conformal) 
gravity multiplet is not self-conjugate under CPT transformations and to build a CPT invariant theory we must use two separate supermultiplets that are exchanged under CPT. 
These two multiplets have a very different character on twistor space. The multiplet containing the positive helicity graviton is described locally by a vector field $V$, whereas the 
multiplet that contains the negative helicity graviton is instead described locally by a 1-form $B$\footnote{More precisely~\cite{Berkovits:2004jj}, in $\N=4$ conformal supergravity 
$V$ represents an element of $H^1(\PT,T_\PT)$ and defines an infinitesimal deformation of the complex structure of a patch of twistor space that preserves the holomorphic 
section $\rD^{3|4}Z$ of $\Ber(\PT)$. The conjugate field $B$ plays a r{\^o}le similar to that of the heterotic $B$-field. Its curvature $H=\d B$ represents an element of 
$H^1(\PT, \Omega^2_{\rm cl})$.}. Correspondingly, in Berkovits' twistor string the two conformal gravity multiplets are represented by the worldsheet vertex operators 
$V^{I}(Z)\,Y_I$ and $B_I(Z)\,\d Z^I$, respectively~\cite{Berkovits:2004jj}. 

Conformal gravity, being a fourth order theory, contains twice as many on-shell degrees of freedom as Einstein gravity.  If we wish to extract the Einstein supergravity 
multiplets from the vertex operators of the Berkovits twistor string, we should require that
\be
	V^I Y_I=(\I^{IJ}\del_J h) \,Y_I
	\qquad\qquad\hbox{and}\qquad\qquad
	B = \phi\,\la Z,\d Z\ra
\label{N=4vertex}
\ee
for some (local) functions $h(Z)$ and $\phi(Z)$ of homogeneities $+2$ and $-2$, respectively. (See~\cite{Penrose:1976jq,Ward:1980am,Mason:2007ct,Mason:2008jy} --- or 
section~\ref{vertex} below --- for further details.) One of the challenges to be overcome in constructing a twistor string for $\N=8$ supergravity is to understand how to unify these 
`vector field' and `one form' vertex operators as part of a single CPT self-conjugate $\N=8$ multiplet. 

Although it is premature to discuss the spectrum of our model at this point, \eqref{Zcomps} already contains a small hint of the solution: the fields $Z$ and $Y$ are unified into a 
single \emph{worldsheet} supermultiplet. Thus, from the perspective of X, there is no fundamental difference between the two types of vertex operator in~\eqref{N=4vertex}.


\subsection{BRST transformations}
\label{BRST}

To complete the specification of our theory, we need to choose a BRST operator. This will be based on the symmetries of X that act trivially on $\Sigma$, as discussed in 
section~\ref{aut}. 

Consider the following three sets of transformations of the component fields. Firstly, 
\be
	\delta_1 Z^I = \eps^a \rho_a^I\, ,\qquad \delta_1\rho^I_{a} = \eps_aY^I\ , \qquad \delta_1 Y^I = 0
\label{trans1}
\ee
with fermionic parameters $\eps^a$, secondly
\be
	\delta_2 Z^I = \frac{1}{2}\kappa_a^{\ a} Z^I\, , \qquad \delta_2\rho^I_{a} = -\kappa_a^{\ b}\rho\, , \qquad \delta_2Y^I = -\frac{1}{2}\kappa^a_{\ a} Y^I
\label{trans2}
\ee
with bosonic parameters $\kappa_a^{\ b}$, and finally
\be
	\delta_3 Z^I = 0\,,\qquad \delta_3\rho^I_{a}  =-\frac{1}{2}\tilde\eps_a Z^I\, , \qquad\delta_3Y^I = \frac{1}{2}\tilde\eps^a\rho_a^I
\label{trans3}
\ee
with fermionic parameters $\tilde\eps_a$.  These transformations represent the actions of a local symmetry of X on the matter multiplet $\cZ$. On a local coordinate patch 
$U\subset\Sigma$, we may take the parameters $(\eps,\kappa,\tilde\eps)$ to be constant. In this case the action S$_1$ is invariant under~\eqref{trans1}-\eqref{trans3} when 
restricted to $U$.  However, the spins and charges of the component fields mean that if we wish to make sense of these transformations globally over $\Sigma$, then $\eps$ and 
$\tilde\eps$ must transform non-trivially on overlaps. Specifically, we must have 
$\eps^a\in\Pi\Omega^0(\Sigma,\L\otimes K_\Sigma^{-1/2}),\  \ \kappa^a_{\ b}\in \Omega^0(\Sigma,\O)$ and 
$\tilde\eps^a\in\Pi\Omega^0(\Sigma,\L^{-1}\otimes K_{\Sigma}^{+1/2})$, so that they fit together to form a supermultiplet in $\Omega^0(\rX,\D)$. In particular, to 
treat~\eqref{trans1}-\eqref{trans3} globally over $\Sigma$, the parameters $\eps$ and $\tilde\eps$ must depend (smoothly) on the worldsheet coordinates. Thus these 
transformations must necessarily be \emph{gauged}.


\subsubsection{The ghost multiplets}
\label{ghosts}

With non-constant parameters, the matter action is not invariant under~\eqref{trans1}-\eqref{trans3}. To remedy this, and to treat the transformations as redundancies, we follow the 
usual procedure of introducing ghosts. Since the above transformations reflect the actions of $\Sigma$-preserving symmetries of $\rX$ as studied in section~\ref{aut}, 
we introduce a ghost multiplet 
\be
	C\in \Pi\Omega^0(\rX,\D)
\label{C}
\ee
in the parity reverse of the parameter multiplet. We declare $C$ to have ghost number $\n_\gh=+1$. As with the matter field, we can expand $C$ in terms of components as
\be
	C^a(z,\theta) = \gamma^a(z) + \theta^b\,{\rm N}^a_{\ b}(z) + \frac{1}{2}\theta^b\theta_b\,\nu^a(z)\ ,
\label{Ccomps}
\ee
where again $a=1,2$. Recalling from section~\ref{X} that $\Pi\D \cong \C^{2}\otimes K_\Sigma^{-1/2}\otimes\L$, we see that the component fields $\gamma^a$ are a pair of 
are $(-\frac{1}{2},0)$-forms of charge $+1$. They are bosonic ghosts for the fermionic parameters $\epsilon^a$ in~\eqref{trans1}.  The bosonic fields $\nu^a$ are likewise 
$(+\frac{1}{2},0)$-forms of charge $-1$ and are ghosts for fermionic parameters $\tilde\epsilon^a$ in~\eqref{trans3}.  Finally, ${\rm N}_a^{\ b}$ are four fermionic ghosts 
corresponding to the GL$(2;\C)$ transformations of the rank 2 bundle $\C^2\otimes\L$. They are scalars on $\Sigma$ of charge 0. We shall often find it convenient to separate this 
GL$(2;\C)$ as GL$(1;\C)\times{\rm SL}(2;\C)$. Accordingly, as in~\eqref{Rsplit} we write
\be
	{\rm N}^a_{\ b} = \frac{1}{2}\delta^a_{\ b}\, {\rm n} + {\rm n}^a_{\ b} 
\label{n}
\ee
where n$^a_{\ b}$ is symmetric and traceless and ${\rm n} \equiv {\rm tr(N)}$. Note that since ${\rm n}$ is the (anticommuting) gauge parameter for the determinant of 
$\C^2\otimes \L$,  the appropriate parameter for gauge transformations of $\L$ itself is ${\rm n}/2$. This explains various factors of $\frac{1}{2}$ that appear in the BRST 
transformations below.

We also introduce an antighost multiplet
\be
	B\in\Pi\Omega^0(\rX,\Ber(\rX)\otimes\D^\vee)
\label{B}
\ee
of $\n_\gh=-1$ that is conjugate to $C$. This may be expanded as
\be
	B_a(z,\theta) = \mu_a(z) + \theta^b\,{\rm M}_{ab}(z) + \frac{1}{2}\theta^b\theta_b\,\beta_a(z)\ .
\label{Bcomps}
\ee
Since $\Pi\Ber(\rX)\otimes\D^\vee \cong \C^2\otimes K_\Sigma^{+1/2}\otimes\L$, we see that the two bosonic fields $\mu_a$ are $(\frac{1}{2},0)$-forms of charge $+1$, the 
fermionic antighosts M$^a_{\ b}$ are uncharged $(1,0)$-forms, and finally $\beta_a$ are a pair of bosonic $(\frac{3}{2},0)$-forms of charge $-1$.  As with the ghost N, we shall 
often separate the antighost M into its GL$(1;\C)$ and SL$(2;\C)$ parts, writing
\be
	{\rm M}_{ab}= \epsilon_{ab}\,{\rm m} +{\rm m}_{ab}
\label{m}
\ee
with m$_{ab}$ symmetric and traceless.

The ghost action is simply
\be
\begin{aligned}
	{\rm S}_2 &= \frac{1}{2\pi}\int_\rX \d^{1|2}z\, B_a\delbar C^a \\
	&=\frac{1}{2\pi}\int_\Sigma \beta_a\delbar\gamma^a + {\rm m}_{ab}\delbar {\rm n}^{ab} + {\rm m}\delbar{\rm n} + \mu_a\delbar\nu^a\ .
\end{aligned}
\label{S2}
\ee
Except for their non-trivial charges under $\L$, the $\beta\gamma$-system is just two copies of the usual $\beta\gamma$ system of the RNS superstring, the MN-system is 
the standard system associated to fixing the GL$(2;\C)$ symmetry of $\D$, and the $\mu\nu$-systems corresponds to fixing supergauge transformations associated to gauginos. 
The non-trivial charges of these ghost fields mean that the gravitinos and gauginos that they fix are also charged. 

The above behaviour is perhaps reminiscent of a GL$(2;\C)$ gauged supergravity on the worldsheet. However, because we are only gauging those symmetries of $\rX$ that act 
trivially on the bosonic Riemann surface $\Sigma$, we are not actually considering worldsheet gravity itself. Correspondingly, our ghosts live only in the subgroup 
$\Pi\Omega^0(\rX,\D)$ of $\Pi\Omega^0(\rX, T\rX)$ and there is no (fundamental) fermionic $bc$-ghost system. It may seem strange to have gravitinos (albeit non-propagating 
ones supplanted by the $\beta\gamma$-system) but no graviton. Usually in supersymmetry, this is not allowed because the structure of the supersymmetry algebra 
$\{Q, Q^\dagger\} = P$ forces us to gauge Poincar{\'e} transformations if we gauge the supersymmetry. In the present case it is possible to have gravitinos without gravitons 
ultimately because the distribution $\D$ is integrable and $\{\D,\D\}\not\supset T\Sigma$.


\subsubsection{The BRST operator}
\label{BRSTQ}

In the presence of the ghosts, the transformations~\eqref{trans1}-\eqref{trans3} are replaced by BRST transformations generated by the operator
\be
	Q = \frac{1}{2}\oint \d^{1|2}z \left( C^a\la \cZ,D_a\cZ\ra+ B_{a}\!\left\{C,C\right\}^a\right)\ ,
\label{Q}
\ee
where the derivative $D_a \equiv \del/\del\theta^a$ and where $\{C,C\}^a = 2C^bD_bC^a$ denotes the anticommutator --- \emph{i.e.}, the graded Lie bracket on the distribution 
$\D$. The integral is to be taken over a real 1-dimensional cycle $\Gamma\subset\Sigma$ as well as over the fermonic directions. Clearly, $Q$ is a fermionic operator of 
$\n_\gh=+1$, and the spins and charges of the fields and measure $\d^{1|2}z$ combine to ensure that $Q$ is a scalar of charge zero under GL$(2;\C)$.  It is also important to 
notice that the BRST operator depends on our choice of infinity twistor $\la\ ,\ \ra$.

Performing the integrals over the anticommuting coordinates $\theta^a$, \eqref{Q} may equivalently be written as\footnote{Our conventions are that $f^{(a}g^{b)}$ is the 
symmetrized product $\frac{1}{2}(f^a g^b + f^bg^a)$. Two-component indices $a,b,...,$ are raised and lowered using the SL$(2)$-invariant anitsymmetric tensor $\epsilon_{ab}$. In 
particular, for bosonic fields such as $\gamma$ and $\nu$, $\gamma^a\nu_a = \epsilon_{ab}\gamma^a\nu^b = -\epsilon_{ba}\nu^b\gamma^a=-\nu^b\gamma_b$. Pairs of 
anticommuting fields would have an extra minus sign. Some care has been taken to ensure the numerical factors in~\eqref{Qcomps}-\eqref{antighostQ} are correct and compatible 
with the numerical factors in the matter and ghost action.} 
\be
\begin{aligned}
	Q&= \oint \gamma^a\la Y,\rho_a\ra + \frac{1}{2}\nu^a\la Z,\rho_a\ra+\frac{{\rm n}}{2}\,\la Y,Z\ra-\frac{1}{2}{\rm n}^{ab}\la\rho_a,\rho_b\ra\\[0.1em]
	 &\qquad+\beta_a(\frac{{\rm n}}{2}\gamma^a+{\rm n}^a_{\ b}\gamma^b)+\mu_a(-\frac{{\rm n}}{2}\nu^a +{\rm n}^a_{\ b}\nu^b) + {\rm m}\gamma^a\nu_a
	 -{\rm m}_{ab}\left({\rm n}^{(a}_{\ \ c}{\rm n}^{b)c} +\gamma^{(a}\nu^{b)}\right)
\label{Qcomps}
\end{aligned}
\ee
in terms of the component fields~\eqref{Zcomps}, \eqref{Ccomps}~\&~\eqref{Bcomps}. When acting on the matter multiplet $\cZ(z,\theta)$, this operator generates the 
transformations
\be
\begin{aligned}
	\delta Z^I &= \gamma^a\rho_a^I + \frac{{\rm n}}{2} Z^I\\[0.1em]
	\delta\rho_a^I &= \gamma_a Y^I +\frac{1}{2}\nu_aZ^I -{\rm n}_a^{\ b}\rho_b^I \\
	\delta Y^I &= \frac{1}{2}\nu^a\rho_a^I-\frac{{\rm n}}{2}Y^I
\end{aligned}
\label{matterQ}
\ee
generalizing~\eqref{trans1}-\eqref{trans3}. Similarly, the BRST transformations act as
\be
\begin{aligned}
	\delta\gamma^a &=  \frac{{\rm n}}{2} \gamma^a+{\rm n}^a_{\ b}\gamma^b \ \qquad\qquad &\delta\nu^a &=  - \frac{{\rm n}}{2}\nu^a+{\rm n}^a_{\ b } \nu^a\\[0.1em]
	\delta{\rm n} &=\nu^a\gamma_a\ \qquad\qquad &\delta{\rm n}^{ab} &= {\rm n}^{(a}_{\ \ c}{\rm n}^{b)c}+\gamma^{(a}\nu^{b)}
\end{aligned}
\label{ghostQ}
\ee
on the ghost multiplet. These transformations directly reflect the structure of the algebra~\eqref{algebra}-\eqref{algebra2}. For example, we see that $\gamma^a$ transform in the 
fundamental of SL$(2;\C)$ and have charge $+1$ under $\L$, while $\nu^a$ is again in the fundamental of SL$(2;\C)$ but has charge $-1$. The unusual factors of $\nu\gamma$ 
in the transformation of n and n$^{ab}$ come from the fact that $\{\v_1,\tilde\v_2\} = R'_{12}$ in~\eqref{algebra}-\eqref{algebra2}. Finally, the BRST transformations act on the 
antighost multiplet as
\be
\begin{aligned}
	\delta\mu_a &= \frac{1}{2}\la \rho_a,Z\ra + \frac{{\rm n}}{2}\mu_a - {\rm n}_a^{\ b}\mu_b  +{\rm m}\gamma_a+{\rm m}_{ab}\gamma^b\\[0.1em]
	\delta{\rm m} &=\frac{1}{2}\left(\la Z,Y\ra -\beta_a\gamma^a +\mu_a\nu^a\right)\\[0.1em]
	\delta{\rm m}_{ab} &=\frac{1}{2} \la \rho_a,\rho_b\ra - 2{\rm n}_{(a}^{\ \ c}{\rm m}_{b)c}^{\phantom{a}} +\beta_{(a}\gamma_{b)}+ \mu_{(a}\nu_{b)}\\
	\delta\beta_a &= \la \rho_a,Y\ra -\frac{1}{2}{\rm n}\beta_a - {\rm n}_a^{\ b}\beta_b -{\rm m}\nu_a +{\rm m}_{ab}\nu^b \ ,
\end{aligned}
\label{antighostQ}
\ee
giving the currents conjugate to each symmetry. The transformations~\eqref{matterQ}-\eqref{antighostQ} have been checked to be nilpotent and to be symmetries of the full 
action ${\rm S}_1+ {\rm S}_2$. Of course, the statement that $Q^2=0$ is subject to potential anomalies --- so too is the very definition of the composite $Q$ operator itself. We shall 
investigate these anomalies in section~\ref{anomalies} below.

BRST invariant configurations may be found by setting the fermionic fields to zero and asking that they remain zero under a BRST transformation. In this regard, the most 
dangerous looking transformation is
\be
	\delta \rho^I_a = \frac{1}{2}\nu_a Z^I
\label{danger}
\ee
which potentially forces $Z$ to vanish, ruining the interpretation of our model as a map to projective twistor space. Even if only some components of the supertwistor $Z^I$ were 
forced to vanish, this would still place intolerable restrictions on the map and destroy any chance of the model describing (non self-dual) gravity. Of course, the resolution is that 
in fact $\nu=0$. To see that this is so, notice that since $\deg\L\geq 0$, the field $\nu^a\in\Omega^0(\Sigma,K_{\Sigma}^{1/2}\otimes\L^{-1})$ 
has no zero-modes (at least generically, and always at $\g=0$). Thus the path integral 
\be
	\int\D\mu\ \exp\left(\frac{1}{2\pi}\int_{\Sigma} \mu_a\delbar\nu^a\right) 
\ee
over the non-zero modes of the conjugate fields $\mu$ imposes the constraint $\nu=0$, rendering~\eqref{danger} harmless. Likewise, the transformation 
$\delta{\rm m}\sim \la Z,Y\ra+\cdots$ is tame because $Y$ vanishes on-shell.

\medskip

To summarize, our model contains just three field multiplets\be
\begin{aligned}
	\cZ\ &\in\ \Omega^0(\rX,\C^{4|\N}\otimes\L)\\
	C\ &\in\ \Pi\Omega^0(\rX, \D)\\
	B\ &\in\ \Pi\Omega^0(\rX, \Ber(\rX)\otimes\D^\vee)\ ,
\end{aligned}
\ee
and is defined by the action\footnote{Recall from~\eqref{Ber} that $\Ber(\rX)\cong\L^2$ and that $\d^{1|2}z$ is a holomorphic section of the trivial bundle 
$\Ber(\rX)\otimes\L^{-2}\cong\O$.}
\be
	{\rm S}=\frac{1}{2\pi}\int_\rX\d^{1|2}z\left(\frac{1}{2}\la\cZ,\delbar\cZ\ra + B_a\delbar C^a\right)
\label{fullact}
\ee
and the BRST charge $Q$ of~\eqref{Q}. This is enough to describe perturbative $\N=8$ supergravity, also allowing for gaugings and conformally flat backgrounds such as 
AdS$_4$.  The only structures involved are the choice of infinity twistor $\la\ ,\,\ra$ and the structure of X as a complex supermanifold. As promised in section~\ref{aut}, zero modes 
of $C$ represent (parity reversed) global automorphisms of X that act trivially on $\Sigma$. Similarly, as in section~\ref{def}, parity reversed zero modes of $B$ are Serre dual to 
$H^1(X,\D)$, the tangent space to the moduli space of $\rX$ as a bundle over $\Sigma$.


\subsection{Worldsheet anomaly cancellation}
\label{anomalies}

The worldsheet theory we have defined is chiral --- the matter and ghost kinetic terms each involve only the worldsheet Dolbeault $\delbar$ operator  --- so it is potentially rife with 
anomalies. We now investigate these, showing that all (local) anomalies cancel when $\N=8$.

\medskip

We first compute the anomalies in the worldsheet gauge theory. Consider first the GL$(1;\C)$ transformations associated to the non-trivial line bundle $\L$. On the two 
dimensional worldsheet, this anomaly is governed by a bubble diagram with the charged chiral fields running around the loop. It is thus determined by the sums of the squares of 
the charges of the fields, weighted by a sign for fermions. The GL(1) charged fields are the $YZ$ system, giving a contribution $\a=(4-\N)$ to the gauge anomaly, the two 
$\beta\gamma$ systems each giving $\a=1$, and the two $\mu\nu$ systems that also contribute $\a=1$ each. Altogether we have
\be
	\a_{\rm GL(1)} = (4-\N) + 2 + 2 = 8-\N
\label{gaugeanom}
\ee
so the GL$(1;\C)$ gauge anomaly vanishes if and only if $\N=8$. 

If the SL$(2;\C)$ bundle has non-trivial second Chern class, there is a further potential gauge anomaly. From the matter fields, only the $\rho\rho$-system transforms non-trivially 
under SL$(2\;C)$, in the antifundamental representation. The bosonic $\beta\gamma$- and $\mu\nu$-ghosts transform in the fundamental (or antifundamental), while the 
fermionic ${\rm m}_{ab}{\rm n}^{ab}$-system is in the adjoint.  The anomaly coefficient is thus
\be
	\a_{\rm SL(2)} = -\frac{1}{2}(4-\N)\,{\rm tr}_{\rm F}(t^k t^k) + 2\,{\rm tr}_{\rm F}(t^kt^k) - {\rm tr}_{\rm adj}(t^kt^k)\ ,
\label{sl2anom1}
\ee
where $t^k$ denote the generators of SL$(2;\C)$, in the representation indicated by the subscript on the trace. (A sum over $k$ is implied.) Note that the $\rho\rho$ contribution 
has a symmetry factor $\frac{1}{2}$ since they are their own antiparticles. Writing ${\rm tr}_{\rm R}(t^kt^k) = C_2({\rm R})\,{\rm dim}({\rm R})$ in terms of the quadratic Casimir of the 
representation, \eqref{sl2anom1} becomes
\be
	\a_{\rm SL(2)} = \frac{\N}{2} \times 2 C_2({\rm F}) - 3 C_2({\rm adj}) = \frac{3}{4}(\N-8)\ ,
\label{sl2anom}
\ee
where we used the SL$(2;\C)$ quadratic Casimirs $C_2({\rm F}) = \frac{3}{4}$ and $C_2({\rm adj}) = 2$. The worldsheet GL$(2;\C)$ gauge theory is thus completely free from local 
anomalies when $\N=8$.

In addition, we can compute the total Virasoro central charge\footnote{Since our theory does not involve worldsheet gravity --- although there is a worldsheet gravitino associated 
to the $\beta\gamma$-system --- the r{\^o}le of this anomaly is not completely clear to me. Its vanishing nonetheless seems significant.}. From the matter fields, the $YZ$-system 
contributes central charge $\c=2(4-\N)$, twice the (complex) bosonic dimension of the non-projective target space, minus twice the fermionic dimension. The $\rho\rho$-system 
are spin $\frac{1}{2}$ fields on the worldsheet of opposite statistics to $Z$. As in the RNS superstring, they contribute a further $(4-\N)$ to the central charge. From the ghosts we 
have two bosonic $\beta\gamma$-systems each contributing $\c=+11$ as in the RNS string, while fermionic ghosts for the gauge system contribute $\c=-2\times {\rm dim}(GL(2))$. 
Finally, because the spin $\frac{1}{2}$ $\mu\nu$-systems are bosonic, they contribute $\c=-1$ each. The total central charge is thus
\be
	\c = 3(4-\N) + 22 - 8 - 2 = 3(8-\N)
\label{centralcharge}
\ee
and vanishes if and only if $\N=8$. There is in addition a potential mixed GL$(1;\C)$-gravitational worldsheet anomaly $\b$ which equals $8-\N$ and again vanishes when 
$\N=8$. Since ${\rm tr_R}(t^k)=0$ for SL$(2;\C)$, there is never any mixed SL(2)-gravitational anomaly.

\medskip

For an alternative (though equivalent) view of things, performing the path integral over the non-zero modes of all fields leads to determinants of $\delbar$-operators. These 
$\delbar$-operators act on sections of bundles as appropriate for the charges and spins of the fields, and the determinants appear in the numerator or denominator according to 
whether the fields are fermionic or bosonic. The resulting chiral determinants are not functions, but form a section of a determinant line bundle over the moduli space of the gauge 
theory (and, in principle, the complex structure of the worldsheet). In order to make sense of the determinants as these moduli vary, we must find a flat connection on this 
determinant line bundle. A natural connection was provided by Quillen~\cite{QuillenLine}. Its curvature $\F$ is computed by the Freed-Bismut 
formula~\cite{Bismut:1986aa,Bismut:1986wr}
\be
	\F = \int_\Sigma {\rm Td}(T\Sigma) \wedge{\rm Ch}(E)\ ,
\label{QuillenF}
\ee
where the bundle $E$ depends on the fields in question. Letting $x$ denote the first Chern class of $T\Sigma$, $y$ denote ${\rm c}_1(\L)$, and $\G$ denote the SL$(2;\C)$ 
bundle, we find
\be
	\F = \frac{\c}{24} \int_{\Sigma} x\wedge x + \frac{\b}{2}\int_{\Sigma} x\wedge y + \frac{\a}{2}\int_\Sigma y\wedge y - \s\int_\Sigma {\rm c}_2(\G)
\label{vanishing}
\ee
where $\a_{\rm GL(1)}$, $\a_{\rm SL(2)}$, $\b$, and $\c$ are the anomaly coefficients computed above.

\medskip

These potential anomalies would also afflict the BRST charge $Q$ in~\eqref{Qcomps}, since it is a composite operator. For example,  the terms proportional to 
the ghost field ${\rm n}$ each contain potential short distance worldsheet singularities. However, the coefficient of this short distance singularity in the combination 
${\rm n}\left(\la Y,Z\ra  -\beta_a\gamma^a + \mu_a\nu^a\right)$ is the same anomaly coefficient $\a = (4-\N) + 2 + 2$ as before, so $Q$ is well-defined when $\N=8$. Similarly, the 
terms proportional to ${\rm n}_{ab}$ are sensitive to any anomaly in the SL$(2;\C)$ gauge symmetry. Finally, there are short distance singularities that potentially obstruct 
$Q^2=0$, so that $Q$ could not be used as a BRST operator. It is left as an exercise to show that these again cancel when $\N=8$.


\subsubsection{Zero modes}
\label{zero}

The absence of gauge anomalies means the path integral over the complete set of \emph{non-zero} modes of all fields is well-defined, providing a section of a determinant line 
bundle over the moduli space whose Quillen connection is flat. We now examine the properties of the \emph{zero} modes of the fields. 

\medskip

Consider first the charged fields in our theory. For a generic chiral $\beta\gamma$-system, with $\gamma$ taking values in some vector bundle $E$, Serre duality and the 
Riemann--Roch theorem give
\be
	\n_{\gamma_0} - \n_{\beta_0} = h^0(\Sigma,E)- h^1(\Sigma,E) = \int_\Sigma {\rm c}_1({\det}\,E)+\frac{1}{2}{\rm c}_1(T\Sigma) \ .
\label{index}
\ee
In the case at hand, writing $\dd$ for the degree of $\L$, we find
\be
	\n_{Z_0}-\n_{Y_0} = 4|\N \times (\dd+1-\g)
\label{YZ0}
\ee
for the bosonic and fermionic components of the $YZ$-system at genus $\g$, 
\be
	\n_{\gamma_0}-\n_{\beta_0} = 2(\dd+2-2\g)
\label{bg0}
\ee
in total for the two $\beta\gamma$-systems in our theory, and a total of
\be
	\n_{\mu_0} - \n_{\nu_0} = 2\dd
\ee
for the two $\mu\nu$-systems\footnote{Here it is the antighost $\mu$ that has zero modes, at least generically.}. 

The first important consequence of this calculation is that the path integral measure $\D(Z,Y,\beta,\gamma,\mu,\nu)_0$ over these zero modes has net charge
\be
	(4-\N)(\dd+1-\g) + 2(\dd+2-2\g) + 2\dd = (8-\N)(\dd+1-\g)\ ,
\label{zeromodecharge}
\ee
where we recall that by Berezin integration, the integral form $\d\theta$ scales oppositely to $\theta$ for a fermion. Thus, when $\N=8$, the zero mode path integral measure
provides a top holomorphic form on the moduli space of the theory.

However, while the total charge of all the zero modes cancels, there are selection rules associated to the zero modes of the individual fields. Generically, when $\d$ is sufficiently 
larger than $\g$ (and always at $\g=0$) the Kodaira vanishing theorem asserts that the above indices are entirely due to the positively charged fields. Let us examine the 
consequences of these selection rules, concentrating on this generic case. 

To begin with, we have $(\dd+1-\g)$ zero modes of $Z^A=\chi^A$ for each $A=1,\ldots,\N$ running over the fermionic directions of twistor space. These fermionic zero modes 
cause the path integral to vanish unless they are saturated by insertions from the vertex operators. Exactly as in the original twistor string~\cite{Witten:2003nn,Berkovits:2004hg}, 
this leads to a relation between the degree of the curve $Z(\Sigma)\subset\CP^{3|\N}$ and the allowed helicity sector represented by the vertex operator insertions. We will find 
that the vertex operators describe an $\N=8$ gravity supermultiplet, in conventions where the positive helicity graviton is at order $(\chi)^0$ and the negative helicity graviton is at 
$(\chi)^8$. Thus we find the usual relation
\be
	\dd =\k + 1+ \g
\label{MHVdegree}
\ee
between the MHV level $\k$ (`number of negative helicity gravitons, minus 2') and the degree of the curve and genus of the worldsheet\footnote{The genus $\sh$ of the 
image curve $Z(\Sigma)$ obeys $\sh \leq \sg$. In particular at MHV level for $\sg=1$, the worldsheet must double cover a twistor line, branched over four points.}. 

Now we consider the zero modes of the bosonic fields. We have $\dd+2-2\g$ zero modes for each of the two components of $\gamma^a$ and $\dd$ zero modes of each of 
the two $\mu_a$s.  Of course, we also have $(\dd+1-\g)$ zero modes for each bosonic twistor component. In the absence of insertions that depend on these zero modes, the 
integrals over $\gamma$ and $\mu$ would diverge, rendering the path integral ill-defined. As in the RNS string, we will find that the vertex operators have $\delta$-function 
support in these fields, giving a meaningful path integral. For now, recall from the introduction that in the flat space limit, an 
$\n$-particle $\g$-loop amplitude with $\n_\pm$ gravitons of each helicity is a monomial of degree $\n_--1+\g$ in the infinity twistor $\la\ ,\, \ra$ as a form,
and $\n_{+}-1+\g$ in the infinity twistor $[\ ,\,]$ as a Poisson structure\footnote{This monomial behaviour of course refers to the amplitude when written in twistor space. It becomes 
obscure on momentum space because the transformation from twistors to momenta itself involves the infinity twistor. Recall also that $\I$ is degenerate in the flat space limit.}. With 
the aid of~\eqref{MHVdegree}, we may rewrite these numbers as
\be
\begin{aligned}
	\n_--1-\g &= \dd \\
	\n_+-1-\g &= \n-(\dd+2-2\g)\ ,
\end{aligned}
\ee
respectively coinciding with the number of zero modes of each component of $\mu$, and $\n$ minus the number of zero modes of each $\gamma$ component.

\medskip

The remaining fields are the worldsheet spinors $\rho$, which generically have no zero modes, and the ghost system associated with gauging the GL$(2;\C)$ transformations. 
These ghosts certainly do have zero modes. However, at $\g=0$ we will content ourselves to treat these by simply `dividing by vol(GL(2))' -- the path integral will lead yield a form 
that is invariant and basic with respect to a natural GL(2) action, and we descend to the moduli space. While this suffices to recover the $\g=0$ scattering amplitudes 
of~\cite{Cachazo:2012kg}, it is really too naive. We discuss this further in section~\ref{discussion}.


\subsection{Vertex operators}
\label{vertex}

In this section we construct the vertex operators representing BRST cohomology classes. These will correspond to a linearized $\N=8$ 
supergravity multiplet. We also construct picture changing operators required to fix zero modes of the bosonic antighosts.


\subsubsection{The $\N=8$ supergravity multiplet}
\label{sugra}

The odd supervector field $\V$ of~\eqref{sectionD} generates a global holomorphic automorphism of X when $\V\in H^0(\rX,\D)$. In the generic case that $\dd \gg \g$, only the 
lowest two components $\v$ and $R$ in the superfield expansion of $\V$ can be globally holomorphic, with
\be
	\v \in H^0(\Sigma,\D) \qquad\qquad\hbox{and}\qquad\qquad R\in H^0(\Sigma,{\rm End}(\C^2\otimes\L))\ .
\ee
The fermionic symmetries of ${\rm X}\to\Sigma$ correspond to zero modes of the ghosts $\gamma$, with each $\gamma^a$ being one of the $\dd+2-2\g$ holomorphic 
sections of $K_\Sigma^{-1/2}\otimes\L$, while the bosonic symmetries correspond to the zero modes of the ghost ${\rm N}^a_{\ b}$ that are constant. To obtain a moduli space 
whose (virtual) dimension is non-negative, and hence to have a well-defined ghost path integral, we must fix these zero modes.

In our case, the odd vector field $\v=\v^a\del/\del\theta^a$ s generates (smooth) translations of along the fibres of ${\rm X}\to\Sigma$. To fix the associated odd automorphisms of X 
we pick points $p_i\in\Sigma$ and demand that the translations act trivially at these points. In the path integral, these translations are represented by the ghosts $\gamma_a$, so 
we can force the translation to be trivial at some $p_i$ by inserting\footnote{The fact that our vertex operators can depend only on $\gamma$, and not on derivatives of $\gamma$, 
is determined by the requirement that the path integral measure over all fields does actually descend to a measure --- or top holomorphic form --- on the moduli space of 
${\rm X}\to\D$. See~\cite{Witten:2012bh} for an explanation in the context of RNS superstrings.}
\be
	\delta^2(\gamma) = \delta(\gamma^1)\,\delta(\gamma^2)
\label{deltagamma}
\ee
at this point. Each such constraint reduces the dimension of the space of automorphisms by 1, so generically we need to pick (at least) $\dd+2-2\g$ points to remove all the global 
odd automorphisms. The resulting $\delta$-functions absorb the $\gamma$ zero modes, rendering the path integral meaningful.

In the usual case of the RNS superstring, the vertex operator would also include a factor of the fermionic $c$ ghost instructing us to quotient the path integral only by those 
diffeomorphisms of the bosonic Riemann surface $\Sigma$ that act trivially at the $p_i$. For us however, since we only quotienting by diffeomorphisms generated by 
sections of $\D\subset T\rX$, there are no $c$ ghosts. If we do not wish our answer to depend on the choice of $p_i\in\Sigma$, we must integrate over them\footnote{See however 
the discussion in section~\ref{SL2}.}. Because $\gamma_a$ has spin $-\frac{1}{2}$ and charge $+1$ under $\L$, the operator $\delta^2(\gamma)$ should be interpreted as 
a (1,0)-form of charge $-2$. So the simplest type of vertex operators are
\be
	\O_h\equiv \int_\Sigma \delta^2(\gamma)\, h(Z)\ ,
\label{NSfix}
\ee
where $h(Z)$ is a (0,1)-form on $\Sigma$ of charge $+2$. This integral is to be taken over $\Sigma$ at $\theta^a=0$. These vertex operators are closely analogous to Neveu-Schwarz vertex operators in the superstring.


The field $h$ is a twistor representative of an $\N=8$ supergravity multiplet, pulled back to $\Sigma$. Writing $Z^I = ({\rZ}^{\rm a}|\chi^A)$ for the bosonic and fermionic 
components of the twistor, we can expand $h$ as
\be
	h({\rZ}|\chi) = h({\rZ}) + \chi^A\psi_{A}({\rZ}) + \frac{1}{2}\chi^A\chi^B a_{AB}({\rZ}) + \cdots + \left(\chi \right)^8 \tilde h(\rZ)
\label{Hcomps}
\ee
where the coefficient of $(\chi)^p$ is a $(0,1)$-form on twistor space of homogeneity $2-p$. Via the linearized Penrose 
transform~\cite{Eastwood:1981jy,WardWells,Penrose:1986ca}, these states correspond to massless fields with one boson of helicity $+2$, 8 fermions of helicity $+\frac{3}{2}$, 
28 gauge fields of helicity $+1$ and so on until we reach the field $\tilde h$ that corresponds to a negative helicity graviton. More precisely, the Penrose transform asserts that an 
on-shell, linearized $\N=8$ supergravity multiplet corresponds to a cohomology class in $H^1(\PT,\O(2))$, of which $h$ is a representative.

\medskip

While $\O_h$ is to be inserted on a fixed section of ${\rm X}\to \Sigma$ (taken to be the zero section $\theta^a=0$), we also have related vertex operators $\hat\O_h$ that are 
integrated over the entire worldsheet supermanifold X. As usual, these are obtained simply by replacing the factor of $\delta^2(\gamma)$ in~\eqref{NSfix} by the integration 
measure $\d^{1|2}z$ on X. We have
\be
\begin{aligned}
	\hat\O_h &\equiv \int_\rX \d^{1|2}z\ h(\cZ)\\
		&=\int_\Sigma  \frac{\del h(Z)}{\del Z^I}Y^I-\frac{1}{2}\epsilon^{ba}\rho^I_a\rho_b^J \frac{\del^2 h(Z)}{\del Z^I\del Z^J}\ .
\end{aligned}
\label{NSint}
\ee
Note that if $I, J$ correspond to bosonic twistor directions, then $\rho^I$ and $\rho^J$ anticommute, and the $\epsilon^{ab}$ symbol ensures that the second term here is
symmetric in $I,J$. Conversely, if $I,J$ are fermionic directions, the expression is naturally antisymmetric in $I,J$. In particular, this means that~\eqref{NSint} is well-defined as a 
composite operator, with no short distance singularities ---  the potential singularity in either the $\rho\rho$-system or the $YZ$-system are each proportional to the (graded) 
antisymmetric infinity twistor $\I$, so the resulting derivatives on $h$ would vanish.

The first term in $\hat\O_h$ also has a natural meaning in twistor space: since $h$ represents an element of $H^1(\PT,\O(2))$, and in canonical quantization of the 
action~\eqref{S1} we have $Y\sim \del/\del Z$, the term $\frac{\del h}{\del Z^I} Y^I$ represents an element of $H^1(\PT,T_{\PT})$. It therefore describes an infinitesimal deformation 
of the complex structure of twistor space. The non-linear Penrose transform~\cite{Penrose:1976jq} asserts that performing a finite deformation of the complex structure of twistor 
space corresponds to turning on self-dual Weyl curvature in space-time. The holomorphic geometry of twistor space determines the conformal structure of space-time, so an 
arbitrary deformation of this complex structure, generated by an arbitrary $V\in H^1(\PT,T_{\PT})$, corresponds to an arbitrary self-dual solution of the Bach equations 
of \emph{conformal} gravity. However, unlike in the original twistor string, the vertex operators we have found here are associated not to arbitrary vector fields, but rather to vector 
fields\footnote{We have used $\I$ to put the vector field index in the natural place. This is harmless when $\I$ is non-degenerate, and in section~\ref{flat} we shall see it happens 
automatically in the flat space limit.} $V^I_{h}\equiv \I^{IJ}\frac{\del h}{\del Z^J}$ that are Hamiltonian\footnote{This is the reason we denote the gravity multiplet by $h$.} with 
respect to the Poisson structure defined by $\I$. Deforming the complex structure by a such Hamiltonian vector field ensures that the holomorphic Poisson structure is preserved, 
and hence the corresponding deformed space-time still has a preferred metric. This metric is a self-dual solution of the vacuum \emph{Einstein} 
equations~\cite{Penrose:1976jq,Ward:1980am,Atiyah:1978wi}. Extending this to the $\N=8$ multiplet gives a BPS solution to the field equations of supergravity.

As usual, the integrated vertex operator~\eqref{NSint} can be added to the original action
\be
\begin{aligned}
 	{\rm S}_1 \to {\rm S}'_1 &= \int_\rX \d^{1|2}z \left(\la\cZ,\delbar\cZ\ra + h(\cZ)\right)\\
	&=\int_\Sigma \la Y,(\delbar Z+V_h)\ra\  + \ \hbox{fermions} \ ,
\end{aligned}
\label{defact}
\ee
describing strings propagating on a background twistor space with deformed complex structure $\delbar\to\delbar+V_h$. In the twistor string framework, deformations that are not 
self-dual are described perturbatively in terms of higher degree maps (worldsheet instantons). 

\medskip

To summarize, freezing all the $\gamma$ zero modes requires us (in the generic case) to include $\dd+2-2\g$ vertex operators of the form $\O_h$, involving the basic twistor 
wavefunction $h$. The remaining ${\n-(\dd+2-2\g)}$ states are represented by integrated vertex operators $\hat\O_h$ that involve not $h$ itself, but rather its associated 
Hamiltonian vector field $V_h$.  In the introduction we saw that the flat space tree-level scattering amplitude involves precisely $\n-(\dd+2-2\g)$ powers of the infinity twistor $[\ ,\ ]$ 
(\emph{i.e.}, the infinity twistor as a degenerate Poisson structure). In particular, at $\g=0$ the tree amplitudes~\eqref{Freddy} are monomials of degree $\n-\d-2$ in $[\ ,\,]$. We now 
understand that this fact has its origin in the odd automorphisms of $X$ as the total space of a fermionic bundle over a fixed $\Sigma$.

\medskip


\subsubsection{Picture changing operators}
\label{pictures}

The worldsheet supermanifold X also has moduli, even for a fixed Riemann surface $\Sigma$, coming from the freedom to deform the distribution 
$\D\subset T\rX$. As in section~\ref{def}, for fixed $\Sigma$ the tangent space to this moduli space is $H^1(\rX,\D)$.  Our twistor string knows about these moduli via the zero 
modes of the antighost multiplet $B$, which live in the parity reversed Serre dual group $\Pi H^0(\rX,\Ber(\rX)\otimes\D^\vee)\cong H^0(\rX,\C^2\otimes K_\Sigma^{1/2}\otimes \L)$. 
In the generic case with $\dd\gg\g$, the only components of the $B$ multiplet to have zero modes are $\mu$ and M.

Consider first the odd moduli space, associated to zero modes of the bosonic $\mu$ antighost in $H^0(\Sigma,\C^2\otimes K_\Sigma^{1/2}\otimes\L)$. To integrate over the odd 
moduli space  we follow the usual procedure of the RNS superstring and insert $2h^0(\Sigma,K_\Sigma^{1/2}\otimes\L)$ `picture changing operators'
\be
	\Upsilon \equiv 2\,\delta^2(\mu)\, S^aS_a\ ,
\label{Upsgeneral}
\ee
where 
\be
	S_a\equiv \frac{1}{2}\la Z,\rho_a\ra + \frac{1}{2}{\rm n}\mu_a - {\rm n}_a^{\ b} \mu_b + {\rm  m}\gamma_a + {\rm m}_{ab}\gamma^b\\
\label{supercurrentdef}
\ee
is the supercurrent obtained by taking the BRST transformation of  $\mu_a$ as in~\eqref{antighostQ}. These insertions correspond to $\delta$-function 
wavefunctions for the gauginos associated to the $\mu\nu$-system, and fix the $\mu_a$ zero modes. See \emph{e.g.} section 3 of~\cite{Witten:2012bh} for a clear discussion of 
picture changing operators and their relation to fixing parity odd moduli.

Unlike the usual picture changing operators of the RNS string, the operator~\eqref{Upsgeneral} involves two copies of these currents because $\D$ has rank two. Because 
$x\,\delta(x)=0$ we can neglect the terms proportional to the $\mu$ antighost in these supercurrents. Similarly, at $\g=0$ when the antighosts ${\rm M}^a_{\ b}$ have no zero 
modes (and there are no ${\rm N}$ insertions with which to contract), we can neglect their contribution to $S^a$. Then the picture changing operator simplifies to become
\be
	\Upsilon \equiv \frac{1}{2}\,\delta^2(\mu)\, \la Z,\rho^a\ra \,\la Z,\rho_a\ra 
\label{PCO}
\ee
When $\g=0$ we need (a minimum of) $\dd$ such $\Upsilon$ insertions.

The previous expression may be thought to be somewhat formal, because composite operator $\la Z,\rho_1\ra\la Z,\rho_2\ra$ has a potential short distance singularity from the 
$\rho\rho$ contraction. To regularize this, we point split the two $\la Z,\rho\ra$ factors and take a limit as they come together. In terms of a local coordinate $z$ on 
$U\subset\Sigma$, one finds
\be
	\lim_{z'\to z} \left\langle \la Z,\rho_1(z')\ra\ \la Z,\rho_2(z)\ra\right\rangle_{\rho\rho} = \lim_{z'\to z} \frac{\sqrt{\d z'}\sqrt{\d z}}{z'-z}\times \la Z(z') ,Z(z)\ra\ ,
\label{pointsplit}
\ee
where the factors of $\sqrt{\d z'}$ and $\sqrt{\d z}$ arise since $\rho$ is a spinor on $\Sigma$. Expanding the holomorphic field $Z$ as 
$Z^I(z') = Z^I(z) + (z'-z) \,\del_zZ^I(z)+ \cdots$, we see that the pole from the $\rho\rho$ propagator is cancelled by a zero from the antisymmetric infinity twistor, leaving us with a 
finite contribution
\be
	\lim_{z'\to z} \left\langle \la Z,\rho_1(z')\ra\ \la Z,\rho_2(z)\ra\right\rangle_{\rho\rho} = -\la Z,\d Z\ra(z)\, .
\label{Phidiag}
\ee
We now define $\Upsilon$ more precisely  as the normal ordered operator
\be
	\Upsilon \equiv \delta^2(\mu) \left(\frac{1}{2}:\!\!\la Z,\rho^a\ra \la Z,\rho_a\ra\!\!:\, -\ \la Z,\d Z\ra\right)
\ee
in which the local contribution of the $\rho\rho$-system is explicitly accounted for. The normal ordering prescription $:\ \ :$ is understood to mean that we do \emph{not} consider 
contractions between the enclosed fields. 

Actually, since the potential short distance singularity cancelled in~\eqref{pointsplit}, we are free to think of $\Upsilon$ as 
in~\eqref{PCO} without normal ordering. We must then remember to include the local contribution~\eqref{Phidiag} when computing correlation functions involving these operators. 
In practice, this approach turns out to be somewhat simpler.

\medskip

Of course, we could have chosen to represent all the external states by the vertex operators $\O_h$ of~\eqref{NSfix}, rather than use any integrated ones $\hat\O_h$. In 
this description, we would quotient the path integral only by translations of $\Sigma$ inside X that act trivially at $\n>\dd+2-2\g$ points. Since we are quotienting by fewer fermionic 
symmetries, the odd dimension of the moduli space increases and we must integrate over this larger odd moduli space. In the language of ghosts, the additional 
insertions of $\O_h$ provide extra factors of $\delta^2(\gamma)$. To compensate for these we should also construct picture changing operators for the $\beta\gamma$-system, 
inserting $\n-(\dd+2-2\g)$ of them so as to provide a top form on the odd moduli space. This is expected to be the correct approach if one wishes to obtain a 
detailed understanding of the compactification of this moduli space~\cite{Witten:2012ga,Witten:2012bh}. It would be interesting to investigate this further.

\medskip

Finally, when $\g\geq1$ we also have zero modes of the fermionic antighosts M$_{ab}$. Insertions of these amount to constructing a top holomorphic form on the (bosonic) moduli 
space of holomorphic GL$(2;\C)$ bundles on $\Sigma$. For the (generic) case that this bundle is \emph{stable} and $\g\geq2$, this moduli space has dimension 
$3(\g-1)+\g$ and has been extensively studied~\cite{Narasimhan,Atiyah:1982fa,Thaddeus:1992sa,Hitchin:1987mz,Hitchin:1990gq,Verlinde:1988sn,Witten:1991we,Axelrod:1989xt}. We discuss it further in section~\ref{SL2}.


\section{Scattering amplitudes in the flat space limit}
\label{amps}

Our prescription for computing $\n$-point worldsheet correlation functions in the $\g=0$ twistor string is to consider the path integral
\be
	\left\langle\prod_{j=1}^{\sdd+2} \,\O_{h_j}\,\prod_{k=\sdd+3}^\sn \hat\O_{h_k}\, \prod_{l=1}^{\sdd} \Upsilon_l\right\rangle =
	\left\langle
	\prod_{j=1}^{\sdd+2}\int_\Sigma\delta^2(\gamma)\, h_j(Z)\prod_{k=\sdd+3}^{\sn} \int_\rX\d^{1|2}z\, h_k(\cZ)\ \prod_{l=1}^{\sdd} \Upsilon_l\right\rangle\ .
\label{corr}
\ee
In this section, we will use this prescription  to recover the flat space tree-level S-matrix of $\N=8$ supergravity in the form obtained in~\cite{Cachazo:2012kg}. To do so, we will 
need to be able to handle correlation functions of $\beta\gamma$-systems involving operators such as $\delta^2(\gamma)$. A clear explanation of how to achieve this was 
recently provided in~\cite{Witten:2012bh} (see especially section 10). For convenience, the relevant points are summarized in appendix~\ref{bg}.


\subsection{A degenerate infinity twistor}
\label{flat}

To compute scattering amplitudes, we must take the limit as the cosmological constant $\Lambda$  tends to zero. In this limit, the rank of the infinity twistor
\be
	\I_{IJ} =  \left(
			\begin{array}{cc|c}
				\Lambda \epsilon_{\dot\alpha\dot\beta} & 0\, & \ 0  \\
				0 & \epsilon^{\alpha\beta} & \ 0 \\
				\hline
				\phantom{0}&&\\ [-1.2em]
				0 & 0\, & \ \sqrt\Lambda \delta_{AB} 
			\end{array}
		\right)
\ee
we have been working with so far becomes non-maximal. In particular, in the flat space limit we must carefully distinguish between the infinity twistor as a form and the infinity 
twistor in its role as a bivector, since
\be
	\I_{IJ}\I^{JK} = \Lambda\,\delta_I^{\ K} \to 0
\label{inverseI}
\ee
and so they are not equivalent. 

If we were to take the flat space limit naively, the matter action~\eqref{S1} would also become degenerate, with the kinetic terms for the $\mu^{\dot\alpha}$ and $\chi^A$ 
components of the supertwistor $Z^I$ vanishing. To avoid this, \emph{before} taking the limit, we relabel the fields as
\be
	Z^I\to Z^I\qquad \rho_1^I\to\rho^I\qquad \rho_2^I\to \I^{IJ}\tilde\rho_{J}\qquad Y^I\to\I^{IJ}Y_J
\label{rescale}
\ee
and include an overall factor of $1/\Lambda$ in the normalization of~\eqref{S1}. In terms of the rescaled fields, the matter action becomes
\be
	{\rm S}_1 = \frac{1}{2\pi}\int_\Sigma Y_I\delbar Z^I + \tilde\rho_{I}\delbar \rho^I
\label{rescaleS}
\ee
and is independent of the cosmological constant. The ghost fields are unchanged. Having ensured the action remains non-degenerate, we can now freely 
take $\Lambda\to0$, setting\footnote{In the presence of an arbitrary gauge coupling g, we are taking the limit $\Lambda\to 0$, $g\sqrt\Lambda \to 0$. See the discussion at the end 
of section~\ref{infinity}.}
\be
	\I^{IJ} =\left(
			\begin{array}{cc|c}
				\epsilon^{\dot\alpha\dot\beta} & 0 \phantom{.}& \ 0  \\
				0 & 0  \phantom{.} & \ 0 \\
				\hline
				0 & 0 \phantom{.} & \ 0 
			\end{array}
		\right)
		\qquad\hbox{and}\qquad
		\I_{IJ} = \left(
			\begin{array}{cc|c}
				0 & 0\, & \ 0  \\
				0 & \epsilon^{\alpha\beta} & \ 0 \\
				\hline
				0 & 0\, & \ 0 
			\end{array}
		\right)\ .
\label{flatinfty}
\ee 
We follow the standard convention that $\la\ ,\,\ra$ denotes contraction by $\I_{IJ}$ with downstairs indices, involving only the $\lambda$ part of $Z$, whereas $[\ ,\,]$ denotes 
contraction by $\I^{IJ}$ with upstairs indices and involves only the derivatives $\del/\del\mu$ tangent to twistor space (or the $\tilde\lambda$s on momentum space).

While the rescaled action is independent of the infinity twistor, the same cannot be said for the BRST operator
\be
	Q_{\rm matter} = \frac{1}{2}\oint \d^{1|2}z \,\la \cZ,D_a\cZ \ra\ .
\ee
As with the action, we first apply the recaling~\eqref{rescale} with a non-degenerate infinity twistor and then take the limit $\Lambda\to 0$. The matter BRST charge 
becomes
\be
\begin{aligned}
	Q_{\rm flat} &= \oint \gamma^1Y_I\rho^I + \gamma^2 [Y,\tilde\rho] + \frac{1}{2}\nu^1\la\rho,Z\ra + \frac{1}{2}\nu^2\tilde\rho_IZ^I \\
	&\qquad\  + \frac{1}{2}{\rm n}\,Y_I Z^I + \frac{1}{2}\left({\rm n}^{12}+{\rm n}^{21}\right)\rho^I\tilde\rho_I + \frac{1}{2}{\rm n}^{11}\la\rho,\rho\ra 
	+ \frac{1}{2}{\rm n}^{22}[\tilde\rho,\tilde\rho] \ ,
\end{aligned}
\label{Qflat}
\ee
and the presence of the degenerate infinity twistor means that not all supertwistor components appear in all terms; for example, $\la\rho,Z\ra = \rho^{\alpha}\lambda_\alpha$ while 
$[Y,\tilde\rho] = Y_{\dot\alpha}\tilde\rho^{\dot\alpha}$.  A somewhat similar BRST operator occurs in Poisson sigma models, see 
\emph{e.g.}~\cite{Cattaneo:2001ys,Bonechi:2007ar}. It would be interesting to explore this connection further. Again, the ghost BRST charge is unaltered. Similarly, in the flat 
space limit the vertex operators become
\be
\begin{aligned}
	\O_h &= \int_\Sigma\delta^2(\gamma)\,h(Z)\\
	\hat\O_h &= \int_{\Sigma} \left[ Y,\frac{\del h}{\del Z}\right] + \left[\tilde\rho, \frac{\del}{\del Z}\left(\rho^I\frac{\del h}{\del Z^I}\right)\right] 
\end{aligned}
\label{Vflat}
\ee
for the external states and
\be
	\Upsilon  = \delta^2(\mu)\,\la \rho,Z\ra\,\tilde\rho_IZ^I
\label{PCOflat}
\ee
for the picture changing operator. As promised, the integrated vertex operator $\hat\O_h$ naturally depends on the Hamiltonian vector field $\left[\frac{\del h}{\del Z},\ \right]$ 
associated to the infinity twistor as a Poisson structure.


\subsection{The worldsheet Hodges matrix}
\label{Hodges}

We are now in position to recover the flat space gravitational S-matrix from the correlator~\eqref{corr} at $\g=0$. In this section we will show that the worldsheet Hodges 
matrix $\HH$ in~\eqref{Hw} arises from the correlation function of the matter vertex operators $\O_h$ and $\hat\O_h$. 

\medskip

Firstly, we notice that the only insertions of $\gamma$ come from the $\delta$-functions in the fixed section vertex operators $\O_h$. These $\delta$-functions serve to fix the 
integrals over the zero modes of $\gamma_a$, representing elements of $H^0(\Sigma,K_\Sigma^{-1/2}\otimes\L)$. For each flavour $\gamma_1$ and $\gamma_2$, we expand 
$\gamma$ as
\be
	\gamma_a(\sigma)  =  \sum_{i=1}^{\sdd+2} \Gamma_{a\,i}\YY_i(\sigma) + \hbox{non-zero modes}\,,
\label{gexpand}
\ee
where the $\YY_i$ form a basis of the zero modes (written in terms of a homogeneous coordinate $\sigma^\ua$ on the $\CP^1$ worldsheet), and where the $\Gamma_a$s are 
$c$-number constants. In~\cite{Witten:2012bh}, it was explained that for each flavour of $\gamma^a$, the insertion of $\delta$-functions leads to
\be
	\left\langle \prod_{j=1}^{\sdd+2} \delta(\gamma(\sigma_j))\right\rangle_{\!\!\beta\gamma}= \frac{1}{\det({\rm Y})}\, .
\label{firstdet}
\ee
See also the discussion in the appendix. Here, Y is the $(\dd+2)\times(\dd+2)$ matrix whose entries are ${\rm Y}_{ij} = \YY_i(\sigma_j)$. At genus zero, a basis of
$H^0(\Sigma, K_\Sigma^{-1/2}\otimes\O(\dd))$ is given by $\YY_i(\sigma) = (\sigma\d\sigma)^{-1/2}\sigma^{\ua_1}\cdots\sigma^{\ua_{\ssd+1}}$. Computing this determinant and 
including both flavours, the path integral over the $\beta\gamma$-system yields
\be
	\left\langle\prod_{j=1}^{\sdd+2} \delta^2(\gamma(\sigma_j))\right\rangle_{\!\!\beta\gamma} 
	=  \frac{1}{|\sigma_1\cdots\sigma_{\sdd+2}|^2}\,\times\,\prod_{j=1}^{\sdd+2}\, (\sigma_j\d\sigma_j) \ ,
\label{Vdm1}
\ee
where $|\sigma_1\cdots\sigma_{\sdd+2}|$ denotes the Vandermonde determinant
\be
	|\sigma_1\cdots\sigma_{\sdd+2}|\  \equiv  \prod_{\substack{i<j\\ i,j\in \{1,\ldots,\sdd+2\}}} (i j)\ .
\label{Vdm}
\ee
This Vandermonde determinant is precisely the denominator factor~\eqref{Vdmintro1} of ${\det}'(\HH)$ in the introduction, here specialized to the case that we remove the first
$\dd+2$ rows and also the last $\dd+2$ columns in computing a minor of the Hodges matrix $\HH$. (That is, we compute the $(\dd+3)^{\rm rd}$ principal minor.)

\medskip

The $(\n-\dd-2)\times(\n-\dd-2)$ minor of $\HH$ itself comes from the remaining part of the matter vertex operators.  As a first step to understanding this, consider the $\n-\dd-2$ 
insertions of $\hat\O_h$ and temporarily neglect the $\left[Y,\frac{\del h}{\del Z}\right]$ terms. The remaining part of $\hat\O_h$ is bilinear in the worldsheet spinors $\rho$ and 
$\tilde\rho$. These are free fields on $\Sigma$. Since they have no zero modes, all $\rho$ and $\tilde\rho$ insertions must be absorbed by 
contracting them pairwise in all possible combinations. Insertions of $\rho$ and $\tilde\rho$ can be found both in $\hat\O_h$ and in the picture changing operators $\Upsilon$. 
However, with our degenerate infinity twistor, $\hat\O_h$ involves only the $\dot\alpha$ components of $\tilde\rho$, while $\Upsilon$ involves only the $\alpha$ components of 
$\rho$. The off-diagonal two point function $\la \tilde\rho_{\dot\alpha}(\sigma)\,\rho^\alpha(\sigma')\ra$ vanishes, so the $\tilde\rho_{\dot\alpha}$s from any $\hat\O_h$ insertion 
can contract only with the $\rho^{\dot\alpha}$s present in some other $\hat\O_h$ insertion.  Furthermore, the pieces 
$\rho_{\alpha}\frac{\del h}{\del\lambda_\alpha} + \rho^A\frac{\del h}{\del\chi^A}$ in $\rho^I\frac{\del h}{\del Z^I}$ in~\eqref{Vflat} may be ignored because there is always at least 
one unpaired $\tilde\rho_{\dot\alpha}$ left over in some $\hat\O_h$ and at least one unpaired $\rho_{\alpha}$ left over in some $\Upsilon$, either of which causes the path 
integral to vanish. We conclude that we can consider the $\rho\tilde\rho$ factors in $\hat\O_h$ independently from those in $\Upsilon$, and that we only need consider the 
$\dot\alpha$ terms in $\hat\O_h$.

This being so, consider the correlator\footnote{In this expression $\mu$ denotes the bosonic twistor component, not the antighost, as should be clear from the indices.}
\be
	C(\sigma_{\sdd+3},\ldots,\sigma_{\sn}) \equiv \left\langle 
	\prod_{k=\sdd+3}^{\sn} \left[\tilde\rho,\frac{\del}{\del Z}\left(\rho^{\dot\beta}\frac{\del h_k}{\del\mu^{\dot\beta}}(\sigma_k)\right)\right] 
	\right\rangle_{\!\!\rho^{\dot\alpha}\tilde\rho_{\dot\alpha}}
	= \ \left\langle \prod_{k=\sdd+3}^{\sn} \tilde\rho^{\dot\alpha}\rho^{\dot\beta}\frac{\del^2 h_k}{\del\mu^{\dot\alpha}\del\mu^{\dot\beta}}(\sigma_k) 
	\right\rangle_{\!\!\rho^{\dot\alpha}\tilde\rho_{\dot\alpha}}
\label{Andrew1}
\ee
coming purely from the $\tilde\rho\rho$-system. In terms of the homogeneous worldsheet coordinates $\sigma^\ua$, the two-point function of the $\rho\tilde\rho$-system is
\be
	\la \rho^{\dot\alpha}(\sigma_i)\,\tilde\rho_{\dot\beta}(\sigma_j)\ra = \delta^{\dot\alpha}_{\ \dot\beta}\,
	\frac{(\sigma_i\d\sigma_i)^{\frac{1}{2}}(\sigma_j\d\sigma_j)^{\frac{1}{2}}}{(ij)}\ .
\label{rhoprop}
\ee
Using this propagator to perform all possible contractions in~\eqref{Andrew1} yields
\be
	C(\sigma_{\sdd+3},\ldots,\sigma_{\sn}) = \left|\HH^{(0)}_{(\sn-\ssd-2)\times(\sn-\sdd-2)}\right| \prod_{k=\sdd+3}^{\sn}  h_k(Z(\sigma_k))\, (\sigma_k\d\sigma_k)\ ,
\label{Bndrew1}
\ee
where $ \left|\HH^{(0)}_{(\sn-\ssd-2)\times(\sn-\sdd-2)}\right|$ is the $(\dd+3)^{\rm rd}$ principal minor of the matrix $\HH^{(0)}$ whose elements are
\be
	\HH^{(0)}_{ij} = \frac{1}{(ij)}\left[\frac{\del}{\del Z_i},\frac{\del}{\del Z_j}\right] = \frac{1}{(ij)}\left[\frac{\del}{\del \mu_i},\frac{\del}{\del \mu_j}\right]
	\qquad \hbox{for $i\neq j$}
\label{Hoffdiag}
\ee
and zero on the diagonal.	 In~\eqref{Hoffdiag} we understand that $\del/\del Z_i$ acts on $ h_i$ in the product~\eqref{Bndrew1}, differentiating this wavefunction with respect to 
(the $\mu^{\dot\alpha}$ component of) its argument $Z(\sigma_i)$. For example, if the external states are taken to be twistor representatives
\be
	 h_i(Z(\sigma_i)) 
	= \int\frac{\d s_i}{s_i^3}\, \bar\delta^2(\lambda_i- s_i\lambda(\sigma_i))\,
	\exp \left(s_i \mu^{\dot\alpha}(\sigma_i)\tilde\lambda_{i\dot\alpha} + s_i\chi^A(\sigma_i)\eta_{iA}\right)
\label{momeig}
\ee
of momentum eigenstates, then $\HH^{(0)}_{ij} \to s_is_j [ij]/(ij)$.

$\HH^{(0)}$ is not quite the full worldsheet Hodges matrix~\eqref{Hw}, because the diagonal elements $\HH^{(0)}_{ii}$ vanish. To recover~\eqref{Hw} in its entirety, we now 
consider the additional effect of the $\left[Y,\frac{\del h}{\del Z}\right]$ terms in the integrated vertex operators~\eqref{Vflat}. With the degenerate infinity twistor, only the 
$\dot\alpha$ component of $Y$ is present here. This $Y_{\dot\alpha}$ cannot contract with any $Z$ in $\Upsilon$, either because there is no short distance singularity or because 
we would again be left with an unpaired $\tilde\rho$. However, the $Y_{\dot\alpha}$ from any given $\hat\O_h$ may contract with the $Z$s in the wavefunctions $h(Z)$ in any of 
the remaining $\hat\O_h$ operators, and also in the `fixed' $\O_h$ vertex operators.

Since  $Y\in \Omega^0(\Sigma,K_\Sigma\otimes\O(-\dd))$ and $Z\in\Omega^0(\Sigma,\O(\dd))$, the $YZ$-propagator is
\be
	\la Y_I(\sigma_i)\,Z^J(\sigma_j)\ra = \delta_I^{\ J} \frac{(\sigma_i\d\sigma_i)}{(ij)}\prod_{r=1}^{\sdd+1} \frac{(a_r j)}{(a_r i)}\ ,
\label{YZprop}
\ee
where the $\sigma_{a_r}$ in the product on the right are arbitrary. This product ensures that both sides have the correct homogeneity. It arises because $\delbar^{-1}\!f$ is 
ill-defined if $f$ is a (0,1)-form of homogeneity $\dd>-1$; we are always free to modify $\delbar^{-1}\!f\to\delbar^{-1}\!f + g$ where $g$ is an arbitrary holomorphic section of 
$\O(\dd)$, since this is annihilated by $\delbar$. We can fix a choice of propagator by specifying $\dd+1$ points at which $\delbar^{-1}\!f$ vanishes. This is the role of the product 
in~\eqref{YZprop}. Of course, any meaningful expression --- such as the Hodges matrix --- is independent of the choice of these points 
(see~\cite{Hodges:2012ym,Cachazo:2012kg,Cachazo:2012pz}).

Using this propagator, with just one $\left[Y,\frac{\del h}{\del Z}\right]$ insertion we have
\be
\begin{aligned}
	&\left\langle \left[Y,\frac{\del h_{\sdd+3}}{\del Z}(\sigma_{\sdd+3})\right] 
	\prod_{j=\sdd+4}^{\sn} \tilde\rho^{\dot\alpha}\rho^{\dot\beta}\frac{\del^2 h_j}{\del\mu^{\dot\alpha}\del\mu^{\dot\beta}}(\sigma_j)  
	\prod_{i=1}^{\sdd+2} h_i(Z(\sigma_j))\right\rangle \\
	&\qquad
	= \,\left|\HH^{(0)}\right|\,
	\left\langle \left[Y,\frac{\del h_{\sdd+3}}{\del Z}(\sigma_{\sdd+3})\right]\,\prod_{i=1}^{\sdd+2} h_i(Z(\sigma_i))\right\rangle \prod_{j=\sdd+4}^{\sn}(\sigma_j\d\sigma_j)\\
	&\qquad
	=\,\left|\HH^{(0)}\right|
	\left\{-\sum_{\substack{k=1\\[0.2em] k \neq \sdd+3}}^\sn \frac{1}{(\dd\!+\!3\, k)}\left[\frac{\del}{\del \mu_{\sdd+3}},\frac{\del}{\del \mu_k}\right] \,\prod_{r=1}^{\sdd+1}
	\frac{(a_r\, k)}{(a_r \, \dd\!+\!3)}\right\} 
	\,\prod_{i=1}^\sn h_i(Z(\sigma_i))\prod_{j=\sdd+3}^{\sn}(\sigma_j\d\sigma_j)
\label{Phitildediag}
\end{aligned}
\ee
where in the first step we integrated out the $\rho$ fields using~\eqref{Bndrew1}. The term in braces in the last line is one of the diagonal elements (the $(\dd+3)^{\rm rd}$ 
diagonal entry) that we were missing from the full Hodges matrix.

We now prove inductively that the sum of $Y$ and $\rho\tilde\rho$ insertions in each integrated vertex operator $\hat\O_h$ means that the total correlation function assembles 
itself into the complete worldsheet Hodges matrix $\HH$. To start, this is certainly true when there are only 2 integrated vertex operators (corresponding to an $\n$-point 
$\overline{\rm MHV}$ amplitude). For
\be
\begin{aligned}
	&\left\langle  \prod_{i=1}^2 \left(\left[Y,\frac{\del h_i}{\del Z}\right] +\tilde\rho^{\dot\alpha}\rho^{\dot\beta} 
	\frac{\del^2 h_i}{\del \mu_i^{\dot\alpha}\del\mu_i^{\dot\beta}}\right)\prod_{j=3}^{\sn} h_j\right\rangle\\
	&\ = \left(\HH_{11}\HH_{22} + \begin{vmatrix} \,0 & \,\HH_{12} \\ \HH_{21} & 0 \end{vmatrix}\right) 
	(\sigma_{1}\d\sigma_{1})(\sigma_2\d\sigma_2)\ 
	\prod_{i=1}^\sn  h_i \  = \ \begin{vmatrix} \,\HH_{11} & \HH_{12} \\ \,\HH_{21} & \HH_{22} \end{vmatrix}\ (\sigma_1\d\sigma_1)(\sigma_2\d\sigma_2)\ \prod_{i=1}^\sn  h_i\ ,
\end{aligned}
\label{initial}
\ee
where the first equality follows because we must either take the $\tilde\rho\rho$ term at both sites or at none\footnote{To lighten the notation, we have temporarily reversed our 
convention and taken the \emph{first} $\n-\dd-2$ vertex operators to be integrated and the \emph{last} $\dd+2$ to be fixed. This corresponds to computing the leading principal 
minor of $\HH$.} (If there is only one $\hat\O_h$ insertion --- which occurs only for the 3-pt $\overline{\rm MHV}$ amplitude --- we are forced to take the $Y$ contribution as 
in~\eqref{Phitildediag}; the $\rho$ fields cannot contribute at all.)

To perform the induction, assume that the worldsheet correlator correctly gives the determinant of the full Hodges matrix when there are $\m-1$ $\hat\O_h$ operators, for some 
value of $\m$. Then from~\eqref{Phitildediag}
\be
	\left\langle \left[Y,\frac{\del h_1}{\del Z_1}\right] \ \prod_{j=2}^{\sm} \hat\O_{h_j} \prod_{k=\sm+1}^\sn h_k\right\rangle
	= \HH_{11} \times  \left|\HH_{(\sm-1)\times(\sm-1)}\right|\times \prod_{i=1}^\sn  h_i\,\prod_{j=1}^{\sm}(\sigma_j\d\sigma_j)\ ,
\label{crumbs}
\ee
where the sum in each of the diagonal elements of the $(\m-1)\times (\m-1)$ minor $\left|\HH_{(\sm-1)\times(\sm-1)}\right|$ also runs over site 1, since the $Y$ insertions in the 
$\hat\O_{h_j}$s leading to this matrix may additionally contract with site 1. The Hodges factors on the right hand side of this expression can be written as the determinant of an 
$\m\times\m$ symmetric matrix with $\HH_{1j}=0$ for $j\neq1$. On the other hand, instead choosing the $\tilde\rho\rho$ term at site 1 gives\footnote{The lines in the matrix 
in~\eqref{crickey} are simply to distinguish contributions from the new insertions at site 1 from the previous inductive step. $\HH$ is \emph{not} a supermatrix.}
\be
	\left\langle \tilde\rho^{\dot\alpha}\rho^{\dot\beta}\frac{\del^2 h_1}{\del Z_1^{\dot\alpha}\del Z_1^{\dot\beta}} 
	\prod_{j=2}^{\sm} \  \hat\O_{h_j} \!\prod_{k=\sm+1}^\sn \!\! h_k\,\right\rangle 
	= \det\left(\begin{array}{c|cccc}
		0 & \HH_{12} & \HH_{13} & \cdots & \HH_{1\sm} \\
		\hline 
		\HH_{12} & \HH_{22} &  \HH_{23} & \cdots  & \HH_{2\sm} \\
		\HH_{13} & \HH_{23} & \ddots & &\\
		\vdots &  \vdots & &   & \vdots \\
		\HH_{1\sm} & \HH_{2\sm} & &  \cdots & \HH_{\sm\sm}
		\end{array}\,\right)
		\prod_{i=1}^\sn  h_i \prod_{j=1}^{\sm}(\sigma_j\d\sigma_j)
\label{crickey}
\ee
where the first row and first column represent the possible choices of contraction for the additional $\tilde\rho$ and $\rho$ insertions. As before, the sum in the diagonal entries of 
the $(\m-1)\times(\m-1)$ Hodges matrix should be extended to run over site 1. It is now clear that~\eqref{crumbs} \&~\eqref{crickey} combine to give the determinant of the full 
Hodges matrix appropriate to $\m$ insertions of $\hat\O_h$ and $\n-\m$ insertions of $\O_h$.

\medskip

Combining this with the factor~\eqref{Vdm1} from fixing the $\gamma$ zero modes shows that the vertex operators~\eqref{Vflat} contribute
\be
	\left\langle\prod_{i=1}^{\sdd+2} \ \O_{ h_i} \! \prod_{j=\sdd+3}^{\sn} \hat\O_{h_j}\right\rangle 
	=\int {\det}'(\HH)\,\prod_{i=1}^{\sn} h_i(Z(\sigma_i))\,(\sigma_i\d\sigma_i)
\label{spiffy}
\ee
to the twistor string path integral, in the specific case that we choose to remove the \emph{same} $\dd+2$ rows and columns in computing a minor of the full $\n\times\n$ 
Hodges matrix (here chosen to be rows and columns 1 through $\dd+2$). In~\cite{Hodges:2012ym,Cachazo:2012kg} we are actually free to compute \emph{any} 
$(\n-\dd-2)\times(\n-\dd-2)$ minor of $\HH$, provided we divide by the corresponding two Vandermonde determinants. We can arrive at these more general representations by 
also allowing `intermediate' vertex operators that involve a single $\delta(\gamma)$ and an integral over the other $\theta$. That is, if we wish to compute a minor of the Hodges 
matrix involving different rows and columns, we should allow
\be
	\int\d\theta_2\,\delta(\gamma_1(\sigma))\left.h(\cZ)\right|_{\theta_1=0}\qquad\qquad\hbox{and}\qquad\qquad
	\int\d\theta_1\,\delta(\gamma_2(\sigma))\left.h(\cZ)\right|_{\theta_2=0}
\label{intermediate}
\ee
as well as $\O_h$ and $\hat\O_h$. The rows and columns that we remove from the Hodges matrix correspond to the independent insertion points of $\delta(\gamma_1)$ and 
$\delta(\gamma_2)$. More generally, it should be clear that since the amplitudes depend on $\HH$ only through ${\det}'(\HH)$,  there is actually a very large amount of freedom in 
the elements themselves. It would be interesting to know if these examples considered in~\cite{Cheung:2012jz} can be realized on the worldsheet. Of course, since ${\det}'(\HH)$ 
is invariant under arbitrary permutations of all $\n$ external states, the minimal case considered in~\eqref{spiffy} is sufficient to recover the amplitude. Indeed, permutation 
invariance is now seen to be a consequence of the usual fact that it does not matter which vertex operators we choose to be `fixed' and which `integrated'. Thus we have 
recovered one of the main ingredients in the formula~\eqref{Freddy} for the tree-level gravitational S-matrix.


\subsubsection{Self-dual $\N=8$ supergravity}
\label{MWact}

In the following section we will show that the remaining, conjugate Hodges matrix comes from the $\dd$ insertions of picture changing operators $\Upsilon$. However, in the 
special case that $\dd=0$ --- corresponding to constant maps to twistor space --- X has no odd moduli and no $\Upsilon$s need be inserted. This case is worth investigating 
separately. 

Since $Z(\sigma)=Z$ for constant maps, we obtain the 3-point $\overline{\rm MHV}$ amplitude
\be
\begin{aligned}
	&\left\langle \int_\Sigma\delta^2(\gamma)\, h_1(Z) \ \int_\Sigma\delta^2(\gamma)\, h_2(Z) \ \int_\rX\d^{1|2}z\,\, h_3(\cZ)\right\rangle\\
	&\ =\int \rD^{3|8}Z\wedge h_1(Z)\wedge \left\{h_2(Z),\,h_3(Z)\right\}\ ,
\label{MHVbar}
\end{aligned}
\ee
where we divided by ${\rm vol(GL(2;\C))}$ in lieu of fixing the zero associated to the worldsheet gauge theory. This absorbs the integration over the three vertex 
operators over $\Sigma$ and ensures the remaining integral is taken over the projective twistor space. As usual, the braces $\{\ ,\,\}$ denote the Poisson bracket associated to the 
infinity twistor $\I$ as a Poisson structure $\I^{IJ}\frac{\del}{\del Z^I}\wedge\frac{\del}{\del Z^J}$. Note that the Poisson bracket itself has homogeneity $-2$, while each $h_i(Z)$ has homogeneity $+2$, so~\eqref{MHVbar} is well-defined on the projective twistor space.

This 3-point $\overline{\rm MHV}$ amplitude is especially important because it is the vertex of the action
\be
	{\rm S_{sd}} = \int\rD^{3|8}Z\wedge\left(h\wedge\delbar h + \frac{2}{3}h\wedge\left\{h,h\right\}\right)\ ,
\label{sdact}
\ee
evaluated on on-shell states. S$_{\rm sd}$ is the twistor action for \emph{self-dual} $\N=8$ supergravity and was first obtained by Mason \& Wolf in~\cite{Mason:2007ct}. At the 
linearized level its equations of motion say that $h$ represents an element of $H^{(0,1)}(\PT,\O(2))$, as we have used in our vertex operators. At the nonlinear level, the equations 
of motion assert that the almost complex structure determined by $\delbar +\{h,\ \}$ is integrable. Once again, the fact that we deform the complex structure only by Hamiltonian vector fields ensures that we have a solution of self-dual Einstein gravity, rather than self-dual conformal gravity~\cite{Penrose:1976jq,Atiyah:1978wi}. S$_{\rm sd}$ is clearly analogous to the holomorphic Chern-Simons theory
\be
	{\rm S_{sdYM}} = \int\rD^{3|4}Z\wedge{\rm tr}\left(A\wedge\delbar A+\frac{2}{3}A\wedge\left[ A,A\right]\right)
\label{sdYMact}
\ee
that describes self-dual $\N=4$ super Yang-Mills in twistor space~\cite{Witten:2003nn}. Just as~\eqref{sdYMact} is the string field theory of the perturbative open B-model, we can 
interpret~\eqref{sdact} as the string field theory of our twistor string, restricted to constant maps.


\subsection{The conjugate Hodges matrix}
\label{cHodges}

When $\dd>0$ we must also account for the picture changing operators $\Upsilon=\delta^2(\mu) \la\rho_1,Z\ra \rho_{2I}Z^I$. These will supply the conjugate Hodges matrix 
$\HH^\vee$.

In the recipe~\eqref{corr} there are no insertions of $\mu$ or $\nu$ except for those in the picture changing operators, which just suffice to absorb the $\mu$ zero-modes. Recall that at $\g=0$ a $\mu$ zero-mode is an element of $H^0(\Sigma,K_\Sigma^{1/2} \otimes\O(\dd))$. Thus, at $\g=0$ we can expand each $\mu_a$ as
\be
	\mu_a(\sigma) =  (\sigma\d\sigma)^{1/2} \ M_{a\,\ua_1\cdots\ua_{\ssd-1}}\sigma^{\ua_1}\cdots\sigma^{\ua_{\ssd-1}} + \hbox{non-zero modes}\ ,
\label{muexp}
\ee
where the $M$s are constants. Following the discussion of the appendix,  we find from~\eqref{fromapp} that the $\delta$-functions associated to freezing the odd moduli produce a factor
\be
	\left\langle\prod_{l=1}^{\sdd}\, \delta^2(\mu(\sigma_l))\,\right\rangle_{\!\!\mu\nu} =\  \frac{1}{\prod_{l=1}^{\sdd} (\sigma_l\d\sigma_l)}\frac{1}{|\sigma_1\cdots\sigma_\sdd|^2}\ .
\label{Vdm2}
\ee
The final factor is the Vandermonde determinant~\eqref{Vdmintro2} appearing in the denominator of ${\det}'(\HH^\vee)$ in~\eqref{Freddy}, specialized to the case that we compute 
the \emph{first} $\dd\times\dd$ minor of $\HH^\vee$. Notice that when $\dd=1$ the unique $\mu$ zero mode is $\mu(\sigma) = (\sigma\d\sigma)^{1/2}$ for each of the two 
antighosts $\mu_1$ and $\mu_2$. Therefore, in the case of the MHV tree, the Vandermonde factor in~\eqref{Vdm2} is replaced by unity. This was the prescription taken 
in~\cite{Cachazo:2012kg}.

\medskip

The numerator of the conjugate Hodges matrix comes from the associated supercurrents.  These are $\la\rho,\lambda\ra\,\tilde\rho_IZ^I 
= \la\rho,\lambda\ra(\tilde\rho_{\dot\alpha}\mu^{\dot\alpha} + \tilde\rho^\alpha\lambda_\alpha+\tilde\rho_A\chi^A)$. Again, the $\tilde\rho\rho$-system has no zero modes, so we 
must absorb all these insertions by contractions. Since $\la\rho,\lambda\ra$ cannot contract with the $\dot\alpha$ or $A$ components of $\tilde\rho$, the only term in the bracket 
that can contribute is $\tilde\rho^\alpha\lambda_\alpha$\footnote{Recall from section~\ref{Hodges} that, with the degenerate infinity twistor of flat space, there could be no 
contribution from cross-contractions of $\tilde\rho$s in $\Upsilon$ with any $\rho$ in $\hat\O_h$.}. Using the propagator 
\be	
	\la\rho_\alpha(\sigma_l)\tilde\rho^\beta(\sigma_m)\ra = \delta_\alpha^{\ \beta}\,\frac{(\sigma_l\d\sigma_l)^{\frac{1}{2}}(\sigma_m\d\sigma_m)^{\frac{1}{2}}}{(lm)}
\ee	
as in~\eqref{rhoprop}, performing all possible $\rho$--$\tilde\rho$ contractions yields
\be
	\left\langle\prod_{l=1}^{\sdd} \la\rho,\lambda\ra\,\tilde\rho^\alpha\lambda_\alpha (\sigma_l) \right\rangle 
	= \left|\HH^\vee_{\ssd\times\ssd}\right| \times\prod_{l=1}^{\sdd}\, (\sigma_l\d\sigma_l)\ ,
\label{detHv}
\ee
where $\left|\HH^\vee_{\ssd\times\ssd}\right|$ is the first $\dd\times\dd$ minor of the matrix with elements
\be
	\HH^\vee_{lm} = \frac{\la\lambda(\sigma_l),\lambda(\sigma_m)\ra}{(lm)} \qquad\hbox{for  $\ l\neq m,\ \ $ and}\qquad
	\HH^\vee_{ll}  = - \frac{\la\lambda(\sigma_l),\d\lambda(\sigma_l)\ra}{(\sigma_l\d\sigma_l)}\ .
\label{Hvelements}
\ee	
To obtain this result, recall that contractions of worldsheet fermions lead to a determinant of a matrix whose $l$-$m^{\rm th}$ entry corresponds to a propagator from site $l$ to 
$m$. The diagonal elements arise as in~\eqref{Phidiag} since we are using the form of picture changing operator without normal ordering, so must allow contractions between 
$\rho$ and $\tilde\rho$ at the same site --- see the discussion in section~\ref{pictures}. Indeed, when $\dd=1$ this is the only contribution.

The off-diagonal elements of $\HH^\vee$ in~\eqref{Hvelements} are exactly the same as those in the conjugate Hodges matrix~\eqref{conjH}. However, the diagonal elements 
in~\eqref{conjH} and~\eqref{Hvelements} appear to be different. Let us now show that~\eqref{conjH} can be simplified so that it takes the form~\eqref{Hvelements}\footnote{I 
am greatly indebted to Lionel Mason for pointing this fact out to me, using a slightly different argument to the one given here.}. To begin with, notice that 
$\la\lambda(\sigma_l),\lambda(\sigma_m)\ra/(lm)$ is everywhere finite, since the potential pole is cancelled by a zero in the numerator. Now consider the diagonal 
term\footnote{Recall that the sum here runs over all $m\in\{1,\ldots,\sn\}$, $m\neq l$. Likewise, the final product is for all $k\in\{1,\ldots,\sn\}$ except $l$ and $m$.}
\be
	(\sigma_l\d\sigma_l) \times \sum_{m\neq l} \frac{\la\lambda(\sigma_l),\lambda(\sigma_m)\ra}{(lm)} \prod_{r=1}^{\sn-\sdd-1} \frac{(a_r m)}{(a_r l)} 
	\prod_{k\neq l,m} \frac{(kl)}{(km)}
\label{Phill}
\ee
of~\eqref{conjH}, where we have included a factor of $(\sigma_l \d\sigma_l)$. The only possible poles in $\sigma_l$ come from the factors 
$(a_r l)$ involving the reference points. However, a key point in~\cite{Cachazo:2012kg,Cachazo:2012pz} was that~\eqref{Phill} was completely independent of these points 
(see~\cite{Cachazo:2012pz} for a contour integral proof of this). Therefore~\eqref{Phill} is actually \emph{holomorphic} in $\sigma_l$. Furthermore, \eqref{Phill} is a scalar of 
homogeneity zero in all other points. Since the first factor in the sum is everywhere finite, the only possible poles in $\sigma_m$ come from the final product and so occur when 
$p_m$ collides with some other marked point $p_k$, with $k,m\neq l$. But for any given $k,m$ (say $m=2$ and $k=3$), it is easy to check that the singularity cancels in the sum. 
Therefore~\eqref{Phill} has no poles in any of the $\sigma_m$ (and hence none in any of the $\sigma_k$). But by Liouville's theorem, a function homogeneous of degree zero that 
is everywhere holomorphic on a compact Riemann surface must be constant. Quite remarkably, we have learnt that~\eqref{Phill} is completely \emph{independent} of all the 
marked points except for $\sigma_i$. Finally, since~\eqref{Phill} is both a $(1,0)$-form in $\sigma_l$ of homogeneity $2\dd$ and is linear in the infinity twistor $\la\ ,\,\ra$, we see 
that
\be
	-\sum_{m\neq l} \frac{\la\lambda(\sigma_l),\lambda(\sigma_m)\ra}{(lm)} \prod_{r=1}^{\sn-\sdd-1} \frac{(a_r l)}{(a_r m)} \prod_{k\neq l,m} \frac{(kl)}{(km)} 
	= -\frac{\la\lambda(\sigma_l),\d\lambda(\sigma_l)\ra}{(\sigma_l\d\sigma_l)}\ .
\ee
This is exactly $\HH^\vee_{ll}$ in~\eqref{Hvelements}. 

With this simplification understood, combining~\eqref{Vdm2} with~\eqref{detHv} shows that the picture changing operators give
\be
	\left\langle \prod_{l=1}^{\sdd} \Upsilon(p_l)\right\rangle = \frac{\left|\HH^\vee_{\ssd\times\ssd}\right|}{|\sigma_1\cdots \sigma_\sdd|^2}
	={\det}'(\HH^\vee)
\label{Hv}
\ee
as the factors of $(\sigma_l\d\sigma_l)$ cancel. This is exactly the contribution of the conjugate worldsheet Hodges matrix in~\eqref{Freddy}, again represented by the specific 
case that we compute the \emph{first} $\dd\times\dd$ minor (the leading principal minor).

\medskip

We now address an issue that may have been puzzling some readers. In the above, we implicitly chose to insert the picture changing operators at $\dd$ of the same points as the 
matter vertex operators. Although this was the minimal choice, was it really necessary? In fact, as in usual superstring theory, the picture changing operators may be inserted at 
completely arbitrary points on the worldsheet, and these locations are not integrated over. Rather than repeat the standard abstract argument for this (for which 
see~\cite{Friedan:1985ge,Polchinski:1998rr}), we shall show directly that despite appearances, \eqref{Hv} is in fact completely independent of the choice of $\dd$ insertion points.

As a warm-up, it is easy to see this claim is certainly true when $\dd=1$, for then  
$\det(\HH^\vee) = \HH^\vee_{11} = -\la\lambda(\sigma_1),\d\lambda(\sigma_1)\ra / (\sigma_1\d\sigma_1)$ and the Vandermonde determinant is unity. Since 
$Z(\sigma)= A\sigma^{\underline 0} + B\sigma^{\underline 1}$ at MHV level, this becomes simply $-\la A,B\ra$ which is obviously independent of the insertion point. 

For the general case, note first that~\eqref{Hv} is homogeneous of degree zero in each of the $\sigma_l$s. The minor of $\HH^\vee$ itself can have no poles --- it is a polynomial in its entries~\eqref{Hvelements}, each of which are everywhere finite. Thus the only possible singularities in~\eqref{Hv} come from the Vandermonde 
determinant $|\sigma_1\cdots\sigma_\sdd|^2$ in the denominator. This produces a second order pole when any pair of insertion points collide. We shall show that this singularity 
is cancelled by a second order zero from $\left|\HH^\vee_{\ssd\times\ssd}\right|$.

To see this, suppose $p_1$ approaches $p_2$ with their separation measured by any small parameter $\eps$ that has a first-order zero when they collide. Then for $m\geq3$ we 
have $\HH^\vee_{1m} \to \HH^\vee_{2m} + O(\eps)$. Subtracting rows and columns, the numerator of~\eqref{Hv} becomes
\be
	\left|\HH^\vee_{\ssd\times\ssd}\right| 
			=\left|\,\begin{matrix}
				{\HH^\vee_{11}}^* & {\HH^\vee_{12}}^* & O(\eps) & & \cdots & & O(\eps) \\
				{\HH^\vee_{12}}^* & \HH^\vee_{22} & \frac{\la 2,3\ra}{(23)} & & \cdots & & \frac{\la 2,\sdd\ra}{(2\sdd)} \\
				O(\eps) & \frac{\la 2,3\ra}{(23)} & \ddots && & & \\
				\vdots & \vdots &  & & & &  \vdots   \\
				O(\eps) & \frac{\la2,\sdd\ra}{(2\sdd)}  & & &  \cdots & & \HH^\vee_{\sdd\sdd}
			\end{matrix}\,\right|
\label{Hvlimit}
\ee
as $\eps$ becomes small. Here, ${\HH^\vee_{12}}^* \equiv \,(\HH^\vee_{12} - \HH^\vee_{22})$. However,  in section~\ref{pictures} $\HH^\vee_{ii}$ was \emph{defined} to be the limit of $\HH_{ij}^\vee$ as the two points collide, so ${\HH^\vee_{12}}^* = O(\eps)$ automatically.  Similarly,
\be
	{\HH^\vee_{11}}^* \equiv  \HH^\vee_{11} -2\HH^\vee_{12} + \HH^\vee_{22}
\ee
is by definition $\eps^2$ times the second derivative of $\HH^\vee_{12}$ at $\sigma_1=\sigma_2$, plus higher order corrections. So ${\HH^\vee_{11}}^*=O(\eps^2)$. Therefore, as 
$p_1\to p_2$ we can extract a factor of $\eps$ from the first row and a separate factor of $\eps$ from the first column in~\eqref{Hvlimit}, showing that the $\dd\times\dd$ minor of 
$\HH^\vee$ indeed has a second order zero in this limit. This cancels the second order pole from the Vandermonde determinant in the denominator so that~\eqref{Hv} remains 
finite. But by the permutation symmetry of $\left|\HH^\vee_{\ssd\times\ssd}\right|$ and $|\sigma_1\cdots\sigma_\sdd|$, \eqref{Hv} cannot have any poles in any of the worldsheet 
coordinates. Again, a function of degree zero that has is globally holomorphic on a compact Riemann surface must be constant, so 
$\left\langle\Upsilon(\sigma_1)\cdots\Upsilon(\sigma_\sdd)\right\rangle$ is completely independent of the insertion points, as expected for picture changing operators.

Above we obtained a representation of ${\det}'(\HH^\vee)$ in which we computed a minor involving the \emph{same} rows and columns. Once again, we can obtain more general 
representations, in which we compute arbitrary minors of $\HH^\vee$, by inserting picture changing operators for the two flavours $(\mu_1,\mu_2)$ of antighost at independent 
locations. That is, we replace
\be
	\Upsilon(\sigma) \to \Upsilon(\sigma,\sigma') \equiv \delta(\mu_1)\la \rho_1,Z\ra(\sigma)\,\times\,\delta(\mu_2)\la\rho_2,Z\ra(\sigma')
\ee
to compute a minor of $\HH^\vee$ from rows and columns corresponding to the independently chosen insertion points of $\delta(\mu_1)$ and $\delta(\mu_2)$. As before, since 
${\det}'(\HH^\vee)$ is competely permutation symmetric in --- indeed, completely independent of --- all insertion points, there is no real difference between any of these cases, 
although a judicious choice may help simplify some calculations.

\medskip

The conjugate Hodges matrix appeared to be the most complicated ingredient in the gravitational scattering matrix as presented in~\cite{Cachazo:2012kg}. Quite remarkably, it 
has turned out to be one of the simplest.


\subsection{The tree-level S-matrix}
\label{smatrix}

Combining the correlation functions~\eqref{spiffy} \& \eqref{Hv} with the remaining integral over the zero modes of the $YZ$ system --- \emph{i.e.}, the space of holomorphic maps 
$Z:\Sigma\to\PT$ --- and dividing by vol(GL$(2;\C)$) to account for the zero modes of the ghosts associated to the worldsheet gauge theory, we have found that
\begin{multline}
	\left\langle\prod_{i=1}^{\sdd+2}\int_\Sigma\delta^2(\gamma)\, h_i(Z)\prod_{j=\sdd+3}^{\sn} \int_\rX\d^{1|2}z\, \H_j(\cZ)\ 
	\prod_{k=1}^{\sdd} \Upsilon_k\right\rangle \\
	= \int \frac{\d^{4(\sdd+1)|8(\sdd+1)}Z}{{\rm vol(GL}(2;\C))}\, {\det}'(\HH)\,{\det}'(\HH^\vee)\,
	\prod_{i=1}^{\sn} \int_\Sigma h_i(Z(\sigma_i))(\sigma_i\d\sigma_i)\ .
\label{success}
\end{multline}
Recalling the genus zero relation $\dd=\k+1$ between the degree of the map and the N$^\sk$MHV level, this correlation function is exactly $\M_{\sn,\sk}$ as defined 
in~\eqref{Freddy}. Summing over all $\dd\geq0$ and allowing all $\n\geq 3$ yields the complete tree-level S-matrix of $\N=8$ supergravity. The ability to reproduce this formula for 
the complete classical S-matrix is a highly non-trivial test of our claim that the worldsheet model proposed in section~\ref{model} does indeed describe $\N=8$ supergravity.


\section{Discussion}
\label{discussion}

We have shown that the worldsheet theory defined by the action~\eqref{fullact} and BRST operator~\eqref{Q} provides a twistor string description of $\N=8$ supergravity. 
The model depends on a choice of infinity twistor, and different choices lead to $\N=8$ supergravity on flat or curved space-times, with the R-symmetry gauged or 
ungauged. We showed that in the flat space limit, $\g=0$ worldsheet correlation functions in this theory generate the complete classical S-matrix of $\N=8$ supergravity, in the 
form discovered in~\cite{Cachazo:2012kg} and proved to be correct in~\cite{Cachazo:2012pz}. By interpreting $\N=8$ supergravity as a twistor string, the present work supplies 
the theoretical framework to explain \emph{why} this form for gravitational scattering amplitudes exists.

The ideas presented here suggest many avenues for further exploration. Let us conclude by discussing some of these.


\subsection{The SL$(2;\C)$ system}
\label{SL2}

The most immediately important issue is to properly understand the r{\^ o}le of the worldsheet gauge theory. In the present paper, our primary concern was to reproduce the tree-
level S-matrix~\eqref{Freddy}. At $\g=0$, the rank 2 bundle $\C^2\otimes\L$ is uniquely determined by degree of $\L$. Consequently, the holomorphic GL$(2;\C)$ bundle has no 
moduli at $\g=0$ and the antighost M has no zero modes. We were able to account for the zero modes of N rather naively, taking the quotient of the zero mode path integral by the 
obvious GL$(2;\C)$ action, or `dividing by ${\rm vol(GL(2))}$'.

At higher genus, holomorphic bundles do have a non-trivial moduli space even for a fixed curve $\Sigma$, and this moduli space has been extensively investigated from many 
points of view in both the mathematics~\cite{Narasimhan,Atiyah:1982fa,Thaddeus:1992sa,Hitchin:1987mz,Hitchin:1990gq} and physics 
literature~\cite{Verlinde:1988sn,Witten:1991we,Axelrod:1989xt}. In particular, when $\g\geq2$ the moduli space $\NN$ of \emph{stable} holomorphic SL$(2;\C)$ bundles has 
dimension $3\g-3$, while that of stable GL$(2;\C)$ bundles has dimension $(3\g-3)+\g$ with the extra $\g$ corresponding to the Picard variety of $\L$. A dense open set of $\NN$ 
may be identified with the Teichm{\"u}ller space of $\Sigma$. We have repeatedly mentioned that our twistor string does not involve worldsheet gravity and so its path integral 
does not automatically include an integral over the moduli space of Riemann surfaces. Nonetheless, it does know about Teichm{\"u}ller space via the moduli space of the rank 
2 gauge bundle associated to $\D$. In fact, there is even a natural isomorphism between $H^1(\Sigma,T\Sigma)$ and $H^1(\NN,T\NN)$ (see \emph{e.g.}~\cite{Narasimhan,Hitchin:1987mz,Axelrod:1989xt}), so that deformations of the complex structure of $\Sigma$ and of the SL(2) bundle are to some extent interchangeable. The 
mechanism by which this is realized in the current context, and the implications for the twistor string, cry out for a better understanding.

A closely related issue is the apparent absence of vertex operators inserted at a fixed point $p\in \rX$, rather than on a fixed section $\Sigma\hookrightarrow\rX$. A proper 
understanding of the SL$(2;\C)$ system should include an operator which creates a puncture on $\Sigma$ to which our vertex operator is attached. Including such operators 
should amount to allowing (parabolically stable~\cite{Seshadri}) holomorphic SL$(2;\C)$ bundles that have simple poles at $p_i\in \Sigma$, such that the monodromy of the 
associated flat connection is in a fixed conjugacy class $\G \subset {\rm SL}(2;\C)$. With $\n$ such punctures, the moduli space of such meromorphic bundles has dimension $3\g-3+\n$. 

As an obvious application, vertex operators associated to punctures on X are likely to be important if one wishes to have a worldsheet description of 
factorization~\cite{Witten:2012ga,Witten:2012bh}. The formula~\eqref{amps} for the gravitational scattering amplitudes was shown to obey the expected factorization properties 
in~\cite{Cachazo:2012pz}. However, the derivation given there was rather involved, because by necessity it dealt with the path integral after integrating out everything but the $Z$ 
zero-modes. By working directly with the vertex operators, one should be able to provide a simpler proof (following the general pattern in string theory), as the terms that may 
become singular in the factorization limit are isolated more cleanly.


\subsection{Higher genus}
\label{loops}

The discussion of section~\ref{SL2} has an immediate corollary that perhaps bears some relation to the debate about whether $\N=8$ supergravity could be 
perturbatively finite~\cite{Bern:2009kd,Bern:2011qn,Bjornsson:2010wm,Kallosh:2011dp,Beisert:2010jx,Bossard:2011tq,Boels:2012sy}. Usually, string theory is 
UV finite because the worldsheet theory is modular invariant. We do not integrate over Teichm{\"u}ller space, but rather over its quotient by the mapping class group. This 
renders harmless any potential divergence as Im$(\tau)\to0$,  and this potentially dangerous region becomes (real) codimension 2 rather than codimension 1. In the theory 
studied here though, worldsheet gravity is replaced by a worldsheet gauge theory. As mentioned above, we expect the path integral to involve an integral over the moduli space  
of stable SL$(2;\C)$ bundles, not the moduli space of curves. Although many aspects of the gauge theory (such as the symplectic form) are invariant under the mapping class 
group,  it is not clear that we should really expect to take the quotient by modular transformations. If not, then the potentially dangerous region Im$(\tau)\to0$ is still present. Of 
course, it is perfectly possible that the integrand still happens to have no singularity here --- and indeed we should expect this at low genus --- but this requires 
calculation\footnote{Another intriguing but very speculative idea would be that the theory allows us to take the quotient by a $\sg$-dependent congruence subgroup of the 
mapping class group that becomes trivial when $\sg$ is greater than some minimum value $\sg_0$, signalling the onset of UV divergences.}. We are unable to
offer the usual string theoretic guarantee that there is simply no place for UV divergences to arise.

\medskip

Whatever the fate of $\N=8$ supergravity at higher loops, the current consensus is that we do not expect any UV divergences when $\g < 7$~\cite{Bern:2009kd,Bern:2011qn,Bjornsson:2010wm,Kallosh:2011dp,Beisert:2010jx,Bossard:2011tq,Boels:2012sy}. What prospect does the twistor string have for computing these 
`intermediate' loop amplitudes? Hopefully, the above discussion has made it clear that we cannot give a proper answer to this question without first understanding the r{\^ o}le 
of the worldsheet gauge theory. Nevertheless it is clear that many properties of these amplitudes are correctly reflected by the worldsheet theory. In particular, the 
zero modes of the $\beta\gamma$- and $\mu\nu$-systems will yield higher loop Hodges matrices that have the correct dependence on the infinity twistors $[\ ,\,]$ and $\la\ ,\,\ra$ 
required by factorization (at least in the generic case with $\dd>2-2\g$; in general we would need to work with $\n$ `fixed' vertex operators and an appropriate number of picture 
changing operators for the $\beta\gamma$-system). The factors of $(ij)$ appearing in these matrices at $\g=0$ naturally generalize to the appropriate Szeg{\"o}  kernels at higher 
genus, while the Vandermonde determinants coming from the correlation function of insertions fixing the zero-modes will involve a basis of holomorphic sections of $\L$ over a 
genus $\g$ curve. All these ingredients can be written in terms of (higher-order) theta functions. See~\cite{Dolan:2007vv} for a related discussion in the context of the original 
twistor string models.

Even if successful, it is doubtful that the twistor string would reproduce even one-loop amplitudes in a form that permits direct comparison with known results in the 
literature~\cite{Bern:1998sv} (though some of the expressions found in~\cite{Bern:2005bb,BjerrumBohr:2005xx,Nasti:2007sr} may be closer). A direct assault on the resulting 
integrals is unlikely to be successful; the integrals over the moduli space of higher degree twistor curves is challenging even at $\g=0$~\cite{Roiban:2004yf}. The most promising 
approach is probably to check that the resulting expressions have all the correct factorization properties.


\subsection{Boundary correlation functions in AdS$_4$}
\label{AdS}

In this paper, we concentrated on taking the flat space limit so as to extract gravitational scattering amplitudes and make contact with the known literature. However, the theory is 
equally capable of describing supergravity or gauged supergravity on AdS backgrounds --- we simply keep the infinity twistor or infinity supertwistor non-degenerate. 

On anti de Sitter space, the natural observables are not scattering amplitudes but rather boundary correlation functions. These are obtained by choosing the external 
wavefunctions to represent bulk--to--boundary propagators, that is, solutions of the free equations of motion on AdS, with a singularity on the conformal boundary. Such 
wavefunctions have a very simple description in twistor space, known in the twistor literature as `elementary states' (see \emph{e.g.}~\cite{Penrose:1975a}). For example,
consider the elementary state
\be
	\phi({\rm Z}) = \frac{1}{\rm A\!\cdot\! Z\ B\!\cdot\! Z} \in H^1(\PT,\O(-2))
\label{elementary}
\ee
representing a scalar field in twistor space. If the line AB is chosen to obey $\la {\rm A,B}\ra=0$, then it lies at infinity. In particular, if $\I$ is the non-degenerate infinity twistor associated to AdS$_4$, then this twistor line represents a point $y$ on the thee dimensional conformal boundary. Using the standard incidence relation $\mu^{\dot\alpha}=x^{\alpha\dot\alpha}\lambda_\alpha$,  the Penrose transform of~\eqref{elementary} appropriate to AdS$_4$ is
\be
\begin{aligned}
	K(x,y) &= \oint \frac{\la {\rm Z ,dZ}\ra}{\rm A\!\cdot\! Z\ B\!\cdot\! Z} 
	= \oint \frac{(1+\Lambda x^2)\la\lambda\d\lambda\ra}{({\rm A}_{\dot\alpha}x^{\dot\alpha\alpha} + {\rm A}^\alpha)\lambda_\alpha\ 
	({\rm B}_{\dot\beta}x^{\dot\beta\beta}+ {\rm B}^\beta)\lambda_\beta}\\
	&\propto \frac{(1+\Lambda x^2)}{(x-y)^2}
\label{bulktoboundary}
\end{aligned}
\ee
where we used the non-degenerate infinity twistor  in the measure $\la{\rm Z,dZ}\ra$. This is the bulk to boundary  propagator for a scalar field, written in the coordinates
where
\be
	\d s^2  = \frac{\d x^\mu \d x_\mu}{(1+\Lambda x^2)^2}
\label{metric}
\ee
is the AdS$_4$ metric and where $(x-y)^2$ is computed using the flat metric.

Using states such as~\eqref{elementary}, it should be possible to use the formalism of this paper to compute arbitrary $\n$-point boundary correlators, again in the form of an 
integral over the moduli space of degree $\dd$ curves in $\CP^3$, at least at $\g=0$. One obvious feature is that the $(\n-\dd-2)\times(\n-\dd-2)$ worldsheet Hodges' matrix and 
the $\dd\times\dd$ conjugate Hodges' matrix will combine into a single $(\n-2)\times(\n-2)$ worldsheet matrix, with the off--block--diagonal terms being proportional to the 
cosmological constant. These terms arise because with a non-degenerate infinity twistor, both the $\rho\rho$- and $YZ$-systems have cross-contractions between the $\O_h$ 
vertex operators representing the external states and the picture changing operators $\Upsilon$. Indeed, one could anticipate this happening. The generalization of Hodges' MHV 
amplitude to the worldsheet Hodges' matrices was deduced~\cite{Cachazo:2012kg} starting from the observation that factorization requires the $\n$-particle N$^{\sdd-1}$MHV flat 
space tree amplitude to contain $\n-\dd-2$ powers of $[\ ,\,]$ and  $\dd$ powers of $\la\ ,\,\ra$ when written in twistor space. But with a non-degenerate infinity twistor these two 
objects are really equivalent.

In the twistor string, as in usual string theory, factorization of scattering amplitudes is closely related to collision of vertex operators 
on the worldsheet~\cite{Vergu:2006np,Skinner:2010cz,Cachazo:2012pz}.  
Factorization of boundary correlators in AdS has been investigated recently in~\cite{Raju:2010by,Raju:2011mp,Fitzpatrick:2011ia,Penedones:2010ue}, where it is shown that (tree-level) Witten diagrams in AdS obey a natural generalization of BCFW recursion. An important observation related to this is that the structure
\be
	\mathcal{M}(Z_1,\ldots,Z_\sn) = \sum\int\rD^{3|8}Z\wedge\frac{\d t}{t}\wedge\mathcal{M}_{\rm L} (Z_1+tZ_\sn,\ldots, Z)\wedge\mathcal{M}_{\rm R}(Z,\ldots,Z_\sn)
\label{BCFW}
\ee
of the BCFW recursion relation in twistor space~\cite{Mason:2009sa,ArkaniHamed:2009si} is completely conformally invariant when expressed in terms of external `twistor eigenstates' 
\be
	 h_i(Z) = \bar\delta^{3|8}(Z,Z_i)\equiv\int\frac{\d s}{s^3}\wedge \bar\delta^{4|8}(Z_i+sZ)\ .
\label{Zeig}
\ee
In particular, the infinity twistor arises only via the three-point functions that seed the recursion relation. We can obtain AdS boundary correlators by integrating~\eqref{BCFW} 
against appropriate boundary elementary states $ h_i(Z_i)$. Thus, in twistor space, the BCFW recursion for Witten diagrams in AdS is exactly the same as BCFW recursion for flat 
space amplitudes. Only the translation back to momentum space (associated to the boundary $\del\overline{\rm AdS}$) and the three-point functions are different.  It would be 
fascinating to relate these observations to the structures of Witten diagrams found in~\cite{Raju:2010by,Raju:2011mp,Fitzpatrick:2011ia,Penedones:2010ue}. Of course we are 
limited to the case that the bulk AdS space is four-dimensional.

Finally,  via analytic continuation to dS, boundary correlators of gravitational modes on AdS may even have cosmological applications~\cite{Maldacena:2011nz}. The ideas 
presented here may provide a way to extend the calculations of~\cite{Maldacena:2011nz} to higher-point functions. The $(\n>3)$-point gravitational wave power spectrum is 
admittedly a rather esoteric cosmological observable!


\subsection{Other issues}
\label{other}

We briefly mention various other issues. 

Firstly, the theory we have presented is purely chiral really provides a top holomorphic form on the moduli space. 
It is this `\emph{scattering form}' that was found in~\cite{Cachazo:2012kg}. To recover the actual scattering amplitudes we must still pick a $4\dd$-dimensional\footnote{This is in 
the case that the wavefunctions are represented in terms of Dolbeault cohomology classes $H^{(0,1)}(\PT,\O(2))$. A description in terms of sheaf cohomology would require us to 
pick a $(4\sdd + \sn)$-dimensional cycle.} real integration cycle on which to integrate this form. When $\dd=1$ and $\g=0$, the moduli space is simply complexified 
space-time and the appropriate integration cycle is just a copy of real Minkowski space. For higher degrees the appropriate contour is less easy to define. One possibility, 
suggested in~\cite{Witten:2003nn} and hardwired into Berkovits' model~\cite{Berkovits:2004hg}, is to pick real structures\footnote{Recall that a real structure is an 
antiholormorphic involution squaring to the identity.} $\tau_1$ on $\Sigma$ and $\tau_2$ on $\CP^3$ and ask that the map is equivariant in the sense that 
$Z\circ\tau_1 = \tau_2\circ Z$.  In ultrahyperbolic space-time signature, these real structures fix an $S^1$ equator on $\Sigma$ at $\g=0$ and an $\mathbb{RP}^3$ real slice of 
twistor space. However, some care is needed in the interpretation of wavefunctions on real twistor space (see \emph{e.g.}~\cite{Mason:2009sa} for a discussion). Other integration 
cycles of interest include those that compute factorization channels of amplitudes, ultimately yielding `leading singularities'. It would be good to know whether the twistor string 
naturally picks a preferred integration cycle for us, or whether this is additional data that must be specified.

In this paper, although we identified the relevant transformations of X that were being gauged, we did not attempt to write down a classical action theory that realized this gauge 
symmetry off-shell. Instead, we moved right away to a gauge fixed model together with its ghosts and BRST symmetry. It would be interesting to construct the unfixed model, 
particularly is this would likely shed further light on the r{\^o}le of the GL(2)-system. Such a model would appear to involve 2 charged gravitinos and 2 charged gauginos in 
addition to the GL(2) gauge fields.

Next, the vertex operators $\O_h$ and $\hat\O_h$ that we obtained are natural analogues of Neveu-Schwarz sector vertex operators in the superstring. It is important to know 
what, if anything, the Ramond sector could be in the present context. Unlike conformal gravity modes in the original twistor string, we would not expect Ramond sector operators to 
be generated at $\g=0$ if they are not present (pairwise) in the external states. If they exist, their r{\^o}le at $\g\geq1$ is clearly important to understand.

We saw in section~\ref{MWact} that, when restricted to constant maps, the string field theory of our model is the twistor action for self-dual $\N=8$ supergravity found 
by~\cite{Mason:2007ct}.  The string field theory of the full model should thus include a further term representing worldsheet instanton contributions. Presumably, only the degree 1 
instantons need be included, as is the case in usual string theory~\cite{Dine:1986zy} and as in the analogous twistor action for $\N=4$ super Yang-Mills~\cite{Boels:2006ir}. 
In our context, these would represent off-shell gravitational MHV vertices. This strongly suggests that despite the difficulties~\cite{Bianchi:2008pu} with Risager recursion for 
gravity~\cite{Risager:2005vk,BjerrumBohr:2005jr} an MHV formalism for gravity exists.  It would clearly be of great interest to find a twistor action for non self-dual gravity. The 
deformed worldsheet action~\eqref{defact} perhaps provides a good starting-point. See~\cite{Mason:2008jy} for an earlier attempt to construct a twistor action for gravity. An 
important step in the right direction has recently been taken in~\cite{Penante:2012wd}.

Last but not least, it would be very interesting to revisit the potential existence of a twistor string for pure $\N=4$ super Yang-Mills in the light of this paper. One approach might be 
to try to understand the meaning of the duality between colour and kinematics~\cite{Bern:2010ue} in a twistor framework. This duality has certainly lead to great progress in the 
computation of multi-loop gravitational amplitudes in momentum space, typically with $\n=4$. The similarity between the twistor action~\eqref{sdact} for self-dual gravity 
and~\eqref{sdYMact} for self-dual Yang-Mills is surely no coincidence. Yang-Mills amplitudes are completely permutation symmetric in the external states provided we include 
their colour factor. Perhaps they also admit a Hodges matrix form.

\bigskip


\acknowledgments
\noindent I am very grateful to N. Arkani-Hamed, P. Goddard, J. Maldacena, L. Mason, H. Verlinde and E. Witten for helpful discussions. I am supported by an IBM Einstein 
Fellowship of the Institute for Advanced Study.


\newpage

\appendix

\section{Some properties of algebraic $\beta\gamma$-systems}
\label{bg}

In this appendix we will compute some correlation functions of operators in $\beta\gamma$ systems that are ingredients in computing the twistor string worldsheet 
correlator~\eqref{corr}. Nothing in this appendix is new --- all (and much more besides) may be found in~\cite{Lechtenfeld:1989wu} and in section 10 of~\cite{Witten:2012bh}, 
which we follow closely.

In constructing the twistor string theory, we imposed no reality conditions on the worldsheet fields (see the discussion in section~\ref{other}). Thus the path integral over these 
fields should be understood as a formal algebraic operation. This is exactly the usual case for Berezin integration of fermionic variables, and so the discussion 
of~\cite{Witten:2012bh} proceeds by relating integrals over bosonic fields to integrals over fermionic fields that are easier to understand. Thus we consider the path integral
over anticommuting fields that we call $b$ and $c$ with action
\be
	{\rm S}_{bc} = \frac{1}{2\pi}\int_\Sigma b\,\delbar c\ .
\label{bcact}
\ee
The result of this path integral depends on the quantum numbers of these fields. Without loss of generality, we can assume that
\be
	c\in\Omega^0(\Sigma,L) \qquad\qquad\hbox{and}\qquad\qquad b\in\Omega^0(\Sigma, K_\Sigma\otimes L^{-1})
\label{bcdef}
\ee
for some line bundle $L$, and we assume the $\delbar$-operator in the action in~\eqref{bcact} acts appropriately on sections of $L$. Then zero modes of $c$ are globally 
holomorphic sections of $L$ while zero modes of $b$ are globally holomorphic sections of $K_\Sigma\otimes L^{-1}$. By Serre duality, this is $H^1(\Sigma,L)$. In the case that 
$L$ is a spin bundle, so that $L^2=K_\Sigma$, (generically) neither field has zero modes and the $bc$ path integral is
\be
	 \int\D(b,c)\,\exp\left(-\frac{1}{2\pi}\int_\Sigma b\,\delbar c\right) = \det(\delbar_{K_\Sigma^{1/2}})\ ,
\label{bcpathint1}
\ee
or in other words the determinant of the Dirac operator on $\Sigma$. As explained in~\cite{AlvarezGaume:1986es,Verlinde:1986kw} this may be written in terms of the Riemann 
theta functions associated to $\Sigma$ and the choice of spin structure. When $\g=0$, we may take it to be a constant. 

For any other choice of $L$, at least one of $b$ or $c$ will have zero modes. By the usual rule $\int \d\theta \,\cdot\, 1 = 0$ of Berezin integration, the path 
integral~\eqref{bcpathint1} vanishes. To obtain a non-vanishing result, we must insert exactly enough fields to absorb the zero modes. For simplicity, let us suppose that $c$ has 
some number $m$ of zero modes, so that we may expand it as
\be
	c(z) = \sum_{i=1}^m c_i \YY_i(z)\ +\ \hbox{non zero-modes}
\ee 
where $c_i$ are anticommuting constants and the $\YY_i$ form a basis of $H^0(\Sigma,L)$ (here written in terms of a local coordinate $z\in U\subset\Sigma$). With $m$ 
insertions of $c$, the path integral becomes
\be
	\int\D(b,c)\,c(z_1)\cdots c(z_m)\,\exp\left(-\frac{1}{2\pi}\int_\Sigma b\,\delbar c\right) 
	= {\det}'(\delbar_{L})\times \int \prod_{i=1}^m \d c_i\ c_0(z_1)\cdots c_0(z_m)
\label{withinsertions}
\ee
where $c_0(z) = \sum_i c_i\YY_i(z)$, and where the determinant is provided by the path integral over the non-zero modes of the $bc$-system. Since the $c$'s anticommute, 
\eqref{withinsertions} must be antisymmetric under the exchange of any pair of insertion points $z_i$ and $z_j$. It must also be holomorphic in all of these insertion points. Thus 
we find
\be
	\int\D(b,c)\,c(z_1)\cdots c(z_m)\,\exp\left(-\frac{1}{2\pi}\int_\Sigma b\,\delbar c\right) 
	= {\det}'(\delbar_{L})\times \det ({\rm Y})\ ,
\label{withins}
\ee
where ${\rm Y}$ is the $m\times m$ matrix with entries ${\rm Y}_{ij} = \YY_i(z_j)$. Equation~\eqref{withins} is the standard result for fermions.

In Berezin integration, if $c(z)$ is fermionic then $\delta(c(z)) = c(z)$. In addition, because ${\rm e}^{\tau} = 1+\tau$ if $\tau^2 = 0$, we can represent $\delta(c(z))$ in integral form 
as
\be
	\delta(c(z)) = \int \d\tau \exp \left(\tau c(z)\right)
\label{intdelta}
\ee
where $\tau$ is an auxiliary anticommuting variable. This is clearly analogous to the usual integral representation of the Dirac $\delta$-function. Following~\cite{Witten:2012bh} 
we thus introduce $m=h^0(\Sigma,L)$ such constant anticommuting variables $(\tau_1,\ldots,\tau_m)$ and let $\hat b$ indicate the collection of fields 
$(b(z); \tau_1,\ldots,\tau_m)$. We also introduce the extended action
\be
	\hat {\rm S}_{\hat b c} = \frac{1}{2\pi}\int_\Sigma b\,\delbar c \ - \ \sum_{i=1}^m \tau_i c(z_i)
\label{extact}
\ee
and the extended path integral measure
\be 
	\D(\hat b,c) = \D(b,c)_{\rm n.z.m}\,\d c_1\cdots \d c_m\, \d\tau_1\cdots\d\tau_m\ ,
\label{extmeas}
\ee
where $\D(b,c)_{\rm n.z.m.}$ is the measure on the infinite dimensional space of non zero-modes. Combining~\eqref{intdelta}-\eqref{extmeas} we see that the path 
integral~\eqref{withins} may be rewritten as
\be
	 {\det}'(\delbar_{L})\times \det ({\rm Y}) = \int\D(b,c)\ {\rm e}^{-{\rm S}_{bc}} \,\prod_{i=1}^m\delta(c(z_i)) = \int\D(\hat{b},c)\ {\rm e}^{-\hat{\rm S}_{\hat b c}}\ \,
\label{extended}
\ee
in terms of the extended set of fields and action. The virtue of thinking about~\eqref{withins} in this way is that we have changed a path integral with insertions into a simple path 
integral over a Gaussian action.

\medskip

It is now straightforward to understand the bosonic case that is actually needed in section~\ref{amps}. Suppose $\beta$ and $\gamma$ are fields on $\Sigma$ with exactly the 
same quantum numbers as $b$ and $c$, except that they are commuting fields. In the case that $L^2=K_\Sigma$, in contrast to~\eqref{bcpathint1} we have
\be
	\int \D(\beta,\gamma)\, \exp\left(-\frac{1}{2\pi}\int_\Sigma \beta\,\delbar\gamma\right) = \frac{1}{\det(\delbar_L)}
\label{betgam1}
\ee
giving the inverse of the determinant, as is familiar from Gaussian integration\footnote{With no reality condition on the $\beta\gamma$-system, this is really a definition of what we 
mean by the Gaussian path integral. See sections 3 \& 10 of~\cite{Witten:2012bh}.}. When $L$ is a more general line bundle such that $\gamma$ has zero modes, the path 
integral diverges (or, without a reality condition, is ill-defined) because of the integration over these zero modes. They can be fixed by inserting $\delta$-function operators, and 
again we represent these in integral form as
\be
	\delta(\gamma(z)) = \int \d t\,\exp\left( t\gamma(z)\right)\ .
\ee
Constructing an extended action and path integral measure as before, but now with commuting variables, we again convert multiple insertions of such $\delta$-function operators 
into a Gaussian integral. We thus find
\be
	\int \D(\beta,\gamma)\ \exp\left(-\frac{1}{2\pi}\int_\Sigma\beta\,\delbar\gamma\right)\,\prod_{i=1}^m \delta(\gamma(z_i))
	= \frac{1}{{\det}'(\delbar_L)\,\det ({\rm Y})}
\label{fromapp}
\ee
where Y is the same matrix of zero modes as before. Notice that with our formal algebraic treatment, there is no modulus sign on the determinants on the right hand side. Notice 
also that if $\gamma(z)$ represents a section of $L$, then $\delta(\gamma(z))$ should transform as a section of $L^{-1}$. Both sides of~\eqref{fromapp} transform as sections of 
$\otimes_i L^{-1}|_{z_i}$.

In the main text, we will be interested in the cases $L = K_\Sigma^{-1/2}\otimes \L$ and $L = K_\Sigma^{1/2}\otimes\L$, where $\L$ is a line bundle of degree $\dd$. In particular, 
when $\g=0$, $\L$ is uniquely determined to be $\O_{\CP^1}(\dd)$. The appropriate zero modes are then
\be
\begin{aligned}
	\YY_i(\sigma) &= \frac{\sigma^{\ua_1}\cdots \sigma^{\ua_{\ssd+1}}}{(\sigma\d\sigma)^{1/2}}\qquad	&\hbox{for }\   K_\Sigma^{-1/2}\otimes\L \\
	\YY_i(\sigma) &= \sigma^{\ua_1}\cdots\sigma^{\ua_{\ssd-1}}\,(\sigma\d\sigma)^{1/2}\qquad 		&\hbox{for }\  K_\Sigma^{+1/2}\otimes\L
\end{aligned}
\label{neededzm}
\ee
where $i$ runs over all possible choices of the indices $\ua_1,\ldots, \ua_{\ssd+1}$ or $\ua_1,\ldots,\ua_{\sdd-1}$, respectively. Inserting these zero modes into Y 
in~\eqref{fromapp} gives equation~\eqref{Vdm1} for the zero modes of each flavour of the worldsheet fields $\gamma^aa$, and~\eqref{Vdm2} for the zero modes of each copy of 
the worldsheet field $\mu_a$. Recall that ${\det}'(\delbar_L)$ is a constant at $\g=0$.

\medskip

Far more information about correlation functions in algebraic $\beta\gamma$ systems can be found in~\cite{Witten:2012bh}.

\newpage

\bibliographystyle{JHEP}
\bibliography{TSbib}

\end{document}